\documentclass[%
aip,jap,
 amsmath,amssymb,floatfix,
 preprint,%
]{revtex4-1}

\usepackage[utf8]{inputenc}
\usepackage[T1]{fontenc}
\usepackage{array}
\usepackage{amsmath,amssymb,amsfonts}
\usepackage{algorithmicx}
\usepackage{mathtools}
\usepackage{algorithm}
\usepackage{algpseudocode}
\usepackage{graphicx}
\usepackage{textcomp}
\usepackage{xcolor}
\usepackage{siunitx}
\usepackage{mathtools}
\DeclarePairedDelimiter\abs{\lvert}{\rvert}%

\def\BibTeX{{\rm B\kern-.05em{\sc i\kern-.025em b}\kern-.08em
    T\kern-.1667em\lower.7ex\hbox{E}\kern-.125emX}}
    
\begin{document}

\title{Spin Wave Based Approximate Computing
}

\author{ \vspace{-0.2cm} Abdulqader Mahmoud}
\email{a.n.n.mahmoud@tudelft.nl}
\affiliation{\vspace{-0.2cm} TU Delft, Computer Engineering Laboratory, Delft, The Netherlands}

\author{Frederic Vanderveken}
\affiliation{\vspace{-0.2cm} Imec, Leuven, Belgium}

\author{Florin Ciubotaru}
\affiliation{\vspace{-0.2cm} Imec, Leuven, Belgium}

\author{Christoph Adelmann}
\affiliation{\vspace{-0.2cm} Imec, Leuven, Belgium}

\author{\vspace{-0.2cm} Said Hamdioui}
\affiliation{\vspace{-0.2cm} TU Delft, Computer Engineering Laboratory, Delft, The Netherlands}

\author{Sorin Cotofana}
\affiliation{\vspace{-0.2cm} TU Delft, Computer Engineering Laboratory, Delft, The Netherlands}

\begin{abstract} \vspace{-0.6cm} By their very nature Spin Waves (SWs) enable the realization of energy efficient circuits as they propagate and interfere within waveguides without consuming noticeable energy. However, SW computing can be even more energy efficient by taking advantage of the approximate computing paradigm as many applications are error-tolerant like multimedia and social media. In this paper we propose an ultra-low energy novel Approximate Full Adder (AFA) and a $2$-bit inputs Multiplier (AMUL). The approximate FA consists of one Majority gate while the approximate MUL is built by means of $3$ AND gates. We validate the correct functionality of our proposal by means of micromagnetic simulations and evaluate the approximate FA figure of merit against state-of-the-art accurate SW, \SI{7}{nm} CMOS, Spin Hall Effect (SHE), Domain Wall Motion (DWM), accurate and approximate \SI{45}{nm} CMOS, Magnetic Tunnel Junction (MTJ), and Spin-CMOS FA implementations. Our results indicate that  AFA consumes $43$\% and $33$\% less energy than state-of-the-art accurate SW and \SI{7}{nm} CMOS FA, respectively, and saves $69$\% and $44$\% when compared with accurate and approximate \SI{45}{nm} CMOS, respectively, and provides a $2$ orders of magnitude energy reduction when compared with accurate SHE, accurate and approximate DWM, MTJ, and Spin-CMOS, counterparts. In addition, it achieves the same error rate as approximate \SI{45}{nm} CMOS and Spin-CMOS FA whereas it exhibits $50$\% less error rate than the approximate DWM FA. Furthermore, it outperforms its contenders in terms of area by saving at least $29$\% chip real-estate. AMUL is evaluated and compared with state-of-the-art accurate SW  and \SI{16}{nm} CMOS accurate and approximate state-of-the-art designs. The evaluation results indicate that it  saves at least $2$x and $5$x energy in comparison with the state-of-the-art SW designs and \SI{16}{nm} CMOS accurate and approximate designs, respectively, and has an average error rate of $10$\%, while the approximate CMOS MUL has an average error rate of $12.5$\%, and requires at least $64$\% less chip real-estate. \end{abstract}

\maketitle

\section{Introduction}
While in the last decades CMOS downscaling has been able to enable high performance computing platforms required to process the information technology revolution induced huge data amount  \cite{data1}, it becomes very difficult to keep the same downscaling pace due to \cite{cmosscaling1}: (i) leakage wall, (ii) reliability wall, and (iii) cost wall. This predicts that Moore's law will come to the end soon and, as a result, researchers have started to explore different technologies (e.g., memristors \cite{mem1,mem2,mem3,mem4}, graphene devices \cite{Yande1,gra1,gra2}, and spintronics \cite{ITRS,spin1,spin2,spin3}) among which Spin Wave (SW) stands apart as one of the most promising due to its \cite{amahmoud2,roadmap,amahmoud1,parallelism, parallelism1,fanout10,fanout11}: (i) Ultra-low energy consumption - SW computing depends on wave interference instead of charge movements. (ii) Acceptable delay. (iii) Highly scalable - SW wavelengths can reach the nanometer range. 

Driven by this potential to build  energy efficient circuits, several SW based logic gates and circuits have been reported  \cite{logic21,logic12,logic11,logic17,fanout, parallelism, parallelism1, fanout10,fanout11, logic2,logic3,logic101, logic1,amahmoud1,memory3}. The Mach-Zehnder interferometer was utilized to build a SW NOT gate, which is considered as the first SW computing device \cite{logic21}. Moreover, XNOR, (N)AND, and (N)OR gates were reported by making use of the Mach-Zehnder interferometer \cite{logic12,logic11,logic17}. Whereas the Mach-Zehnder interferometer utilise  SW amplitude to perform the logic operations, other devices utilize SW phase or both  phase and amplitude to build fanout enabled  Majority, (N)AND, (N)OR, and X(N)OR gates \cite{fanout, fanout10,fanout11}. Moreover, SW frequency was utilised as an additional parameter to improve data storage and computing capabilities of multi-frequency Majority and X(N)OR gates \cite{parallelism,parallelism1}. In addition, physical realization of Majority gates were demonstrated \cite{logic2,logic3,logic101}. Furthermore, SW circuits were proposed at conceptual level, i.e., without simulation or experimental results,  \cite{logic1}, at simulation level, $2$-bit inputs SW multiplier \cite{amahmoud1} and magnonic half-adder \cite{wang2020magnonic}, as well as simulation based practical  $mm$ range prototypes \cite{memory3}. 

All the aforementioned logic gates and circuits were designed to provide accurate results, whereas many current applications like multimedia processing and social media are error tolerant and, within certain bounds,  are not fundamentally perturbed by computation errors \cite{applications}. Therefore, such applications can benefit from approximate computing circuits, which can save significant amounts of energy, delay, and area, while providing acceptable accuracy.  In view of this, this paper introduces novel  energy efficient Approximate SW-based Full Adder (AFA) and Approximate $2$-bit inputs Multiplier (AMUL), and its  main contributions can be summarized as follows: 
\begin{itemize}
  \item Developing and designing a SW based approximate FA: The proposed adder consists of one Majority gate and has a $25$ \% error rate.
  \item Developing and designing a SW based Approximate $2$-bit inputs MUL: The proposed AMUL is implemented using $3$ AND gates and has a $10$ \% error rate.
  \item Validation of the proposed AFA and AMUL circuits by means of the MuMax3 software. 
  \item Demonstrating the superiority: The proposed approximate circuits performance is assessed and compared with accurate and approximate state-of-the-art design counterparts. Our results indicate that AFA consumes $43$\% and $33$\% less energy than accurate state-of-the-art SW and \SI{7}{nm} CMOS counterparts, respectively, and saves $69$\% and $44$\% in comparison with accurate and approximate \SI{45}{nm} CMOS, respectively. In addition, it saves more than $2$ orders of magnitude in terms of energy when compared with accurate Spin Hall Effect (SHE) and Domain Wall Motion (DWM), accurate and approximate Magnetic Tunnel Junction (MTJ), and Spin-CMOS based counterparts. In addition, it achieves the same error rate as approximate \SI{45}{nm} CMOS and Spin-CMOS FAs and $50$\% less error rate than the approximate DWM. Also, it requires at least $29$\% less chip real-estate in comparison with the other state-of-the-art designs. Moreover, AMUL saves at least $2$x and $5$x energy in comparison with accurate SW and \SI{16}{nm} CMOS accurate/approximate designs, respectively,  has an average error rate of $10$\%, while the approximate CMOS MUL has an average error rate of $12.5$\%, and requires at least $64$\% less chip real-estate.
\end{itemize}

The paper is organized as follows. Section \ref{sec:Basics of spin-wave technology} provides SW computing background. Section \ref{sec:Proposed approximate functions} introduces the proposed approximate circuits. Section \ref{sec:Simulation Setup and Results} presents the simulation setup and simulation results. Section \ref{sec:Performance Evaluation} provides performance evaluation data and discusses variability and thermal noise effects and Section \ref{sec:Conclusion} concludes the paper. 

\section{Spin Wave Based Technology Basics}
\label{sec:Basics of spin-wave technology}
We explain the SW basics and computing paradigm in this section. 

\subsection{Spin Wave Fundamentals}
\label{sec:spin-wave fundamentals}
The Landau-Lifshitz-Gilbert (LLG) describes the magnetization dynamics caused by the magnetic torque when magnetic material magnetization is out of equilibrium \cite{amahmoud2}

\begin{equation} \label{eq:1}
\frac{d\vec{M}}{dt} =-\abs{\gamma} \mu_0 \left (\vec{M} \times \vec{H}_{eff} \right ) + \frac{\alpha}{M_s} \left (\vec{M} \times \frac{d\vec{M}}{dt}\right ),
\end{equation}
where $\gamma$ is the gyromagnetic ratio, $\alpha$ the damping factor, $M$ the magnetization, $M_s$ the saturation magnetization, and $H_{eff}$ the effective field which contains the different magnetic interactions. In this work, the effective field is the summation of the external field, the exchange field, the demagnetizing field, and the magneto-crystalline field.

For small magnetic perturbations, Equation (\ref{eq:1}) can be linearized and results in wave-like solutions which are known as Spin Waves (SWs), which can also be seen as collective excitations of the magnetization within the magnetic material. Just like any other wave, a SW is completely described by its amplitude $A$, phase $\phi$, frequency $f$, wavelength $\lambda$ , and wavenumber  $k=\frac{2\pi}{\lambda}$. The relation between frequency $f$ and wavenumber $k$ is called the dispersion relation and is very important for the design of the magnonic devices  \cite{amahmoud2}.

\begin{figure}[t]
\centering
  \includegraphics[width=\linewidth]{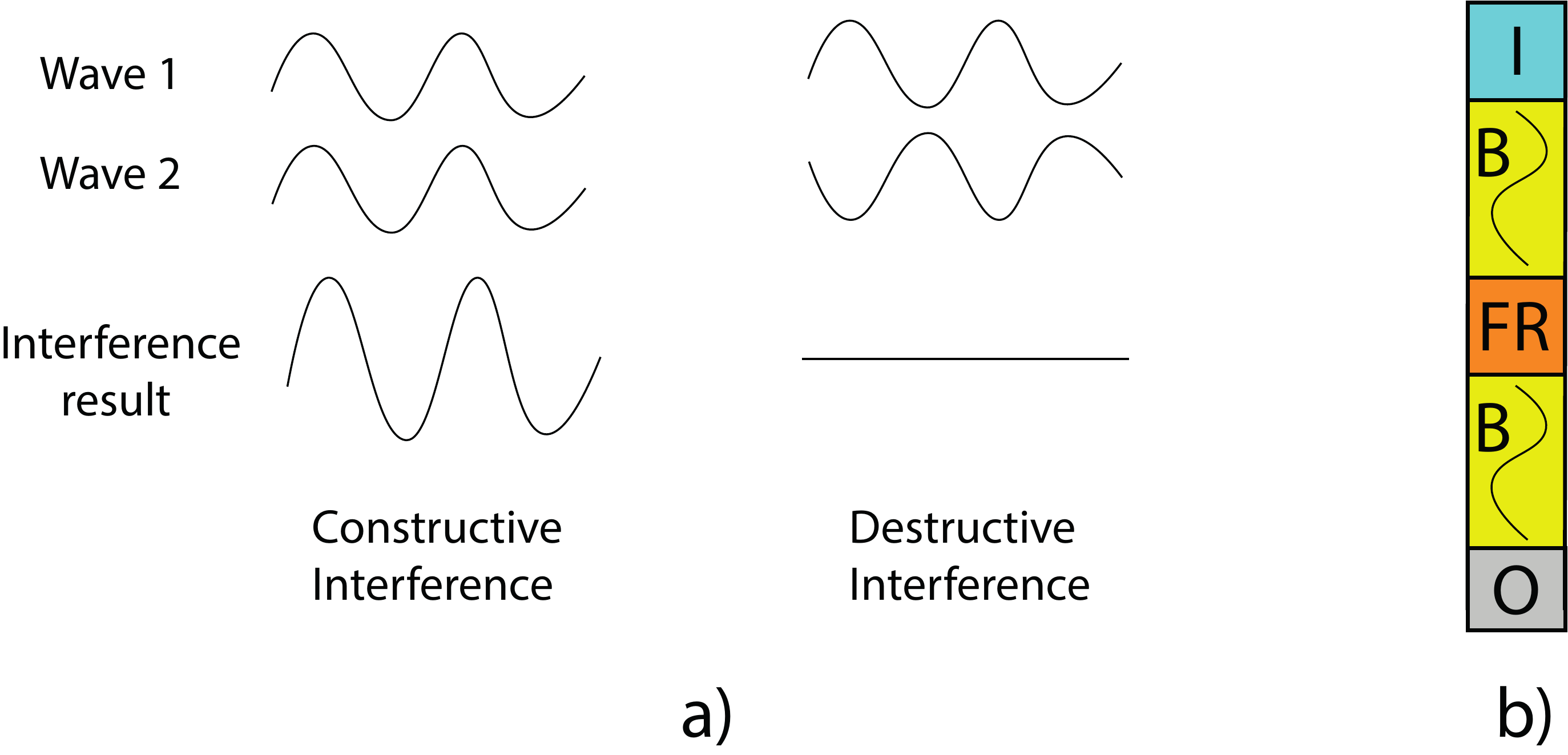}
  \caption{a) Constructive and Destructive Interference. b) Spin Wave Device}
  \label{fig:spin_wave_device}
\end{figure} 

\subsection{SW Computation Paradigm}
\label{sec:SW Computation Paradigm}

The SW amplitude and phase can be used to encode information at different frequencies, which enables parallelism \cite{amahmoud2,parallelism}. The interaction between multiple SWs present in the same waveguide is based on the interference principle. Figure \ref{fig:spin_wave_device}a)  presents an example of interaction between $2$ SWs  excited with the same $A$, $\lambda$, and $f$ in the same waveguide. If the $2$ SWs have the same phase $\Delta \phi=0$, they interfere constructively resulting in a SW with higher amplitude, whereas if they are out of phase $\Delta \phi=\pi$, they interfere destructively, resulting in approximately zero amplitude SW. Moreover, SWs interference provides natural support for Majority function evaluation as if an odd number of SWs interfere the resultant SW is obtained by a Majority decision. For example, if $3$  same $A$, $\lambda$, and $f$ SWs interfere the resultant SW has a phase of $0$ if at most $1$ SW has a phase of $\pi$ and a phase of $\pi$ if at most $1$ SW has a phase of $0$. Note that such a $3$-input Majority CMOS implementation  requires $18$ transistors whereas in SW technology it is implemented using one waveguide only. More complex interference cases exist if the propagating SWs have different $A$, $\lambda$, and $f$, which might be of interest for designing novel magnonic computing systems. However, in this paper, we focus on the simplest case where all the excited SWs have the same $A$, $\lambda$, and $f$ and can take two discrete phases $\phi=0$ and $\phi=\pi$. Logic $0$ refers to a SW with $\phi=0$, and a logic $1$ refers to a SW with $\phi=\pi$.

Figure \ref{fig:spin_wave_device}b) presents a generic SW logic device that consists of four regions: Excitation Stage $I$, Waveguide $B$, Functional Region $FR$, and Detection Stage $O$ \cite{amahmoud2}. In $I$ SWs are generated by means of, e.g., microstrip antennas \cite{amahmoud2}, magnetoelectric cells \cite{amahmoud2}, Spin Orbit Torque \cite{amahmoud2}. $B$ is the medium for SW propagation and can be made of different magnetic materials, e.g., Permalloy, Yttrium iron garnet, CoFeB \cite{amahmoud2}. The waveguide material is an important parameter as it fundamentally  determines the SW properties. In $FR$ SWs can be amplified, normalized or interfere with other SWs. In $O$ the output SW is captured and converted to the electrical domain using the same type of cells as in $I$. 
Two main SW detection techniques are in place \cite{amahmoud2}: phase and threshold based. In phase detection, the output is determined by comparing the detected SW phase with a predefined phase. For example, if the detected SW has a phase of $0$/$\pi$ the output is logic $0$/$1$, respectively. 
Threshold detection determines the output by comparing the detected SW amplitude with a predefined threshold. For instance, if the detected SW amplitude is larger than the predefined threshold, the output is logic $1$ and logic $0$ otherwise. 

\begin{figure}[t]
\centering
  \includegraphics[width=0.7\linewidth]{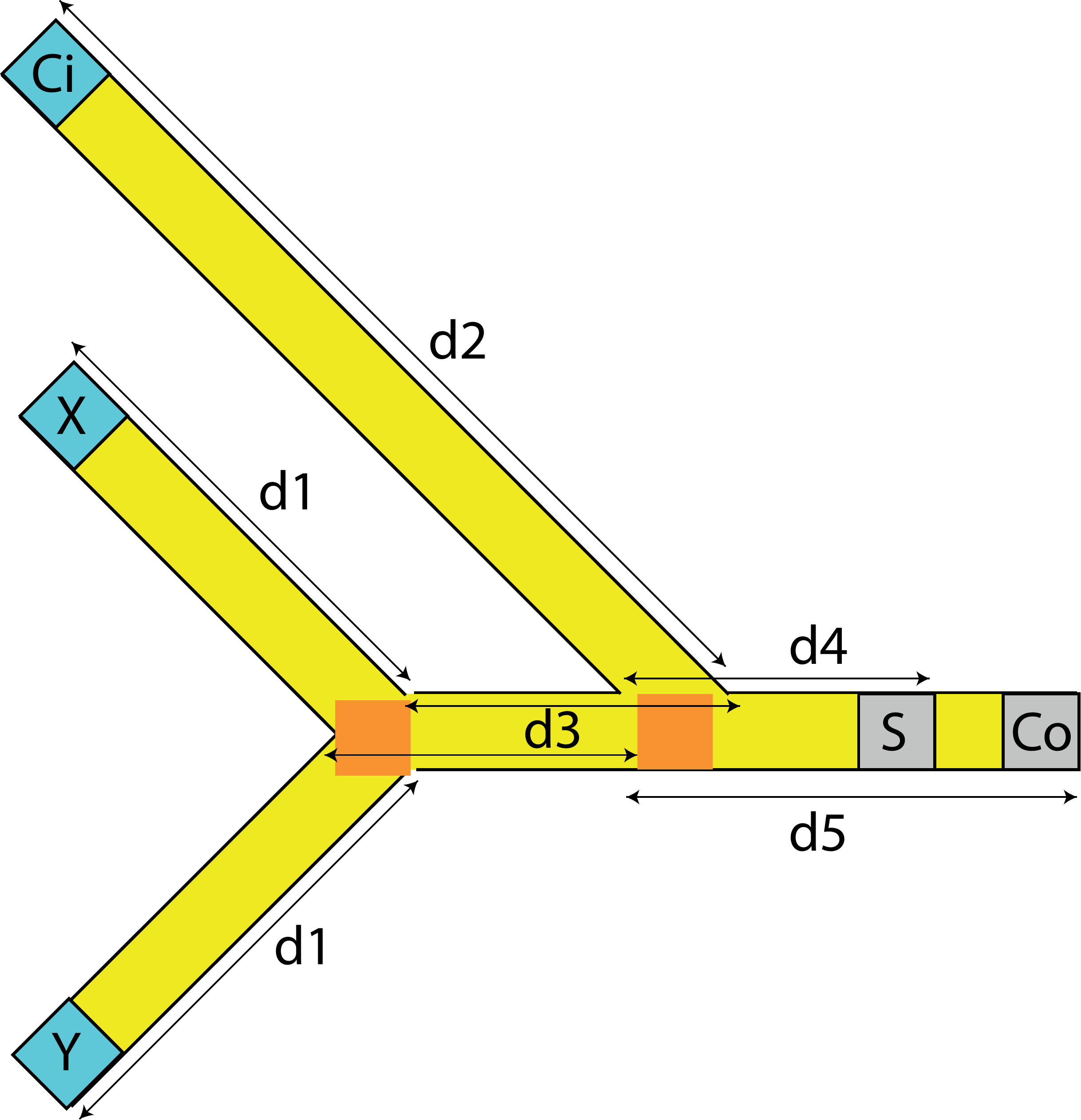}
  \caption{Approximate Spin Wave Based FA.}
  \label{fig:structure1}
\end{figure}

\begin{table}[t]
\caption{Accurate and Approximate SW-based FA}
\label{table:1}
\normalsize
\centering
  \begin{tabular}{|c|c|c|c|c|c|c|c|}
    \hline
   $XYC_i$ & $C_o$ & $S_{ac}$ & $S_{ap}$   \\
    \hline
    $0$ $0$ $0$ & $0$ & $0$ & \textbf{\underline{1}} \\
    \hline
    $0$  $0$  $1$ & $0$ & $1$ & $1$  \\
    \hline
    $0$   $1$  $0$ & $0$ & $1$ & $1$  \\
    \hline
    $0$   $1$  $1$ & $1$ & $0$ & $0$  \\
    \hline
    $1$  $0$  $0$ & $0$ & $1$ & $1$  \\
    \hline
    $1$ $0$  $1$ & $1$ & $0$ & $0$ \\
    \hline
    $1$  $1$  $0$ & $1$ & $0$ & $0$  \\
    \hline
    $1$  $1$  $1$ & $1$ & $1$ & \textbf{\underline{0}} \\
    \hline

    \end{tabular}
\end{table}

\section{SW Approximate Functions}
\label{sec:Proposed approximate functions}

In this section, we introduce and analyse SW-based Approximate Full Adder (AFA) and $2$-bit inputs Multiplier (AMUL).

\subsection{SW Approximate Full Adder}
Figure \ref{fig:structure1} presents the proposed Approximate FA (AFA) structure, which has $3$ inputs $X$, $Y$, and $C_i$, and $2$ outputs $S$ and $C_o$ and is a $3$-input Majority gate that evaluates $S=\overline{C_o}=\overline{MAJ(X,Y,C_i)}$ as suggested in \cite{SPIN}.  AFA generates $C_o$ without any error as it is detected as the Majority of $X$, $Y$, and $C_i$, which is also the case in accurate FAs. On the other hand, $S$ is detected with a $25$\% error rate as $S=\overline{MAJ(X,Y,C_i)}$ approximate the accurate FA Sum, which equals to $S=XOR(XOR(X,Y),C_i)$. 
Table I presents FA and AFA truth tables, which clarifies that the approximate FA sum $S_{ap}$ is erroneous when all inputs are $0$/$1$.

To achieve the AFA behaviour the design in Figure \ref{fig:structure1} has to be properly dimensioned. 
The waveguide width must be smaller or equal to the SW wavelength $\lambda$ and SW amplitude, wavelength, and frequency must be the same at every excitation cell. Furthermore, the structure dimensions must be precisely determined because the  interference pattern depends on the location and distances between different excitation and detection cells. For example, if the constructive  interference pattern is desired when the SWs have the same phase $\Delta \phi=0$ and destructive when the SWs are out-of-phase $\Delta \phi=\pi$, $d_1$, $d_2$, and $d_3$ must be equal with $n\lambda$ (where $n=0,1,2,3,\ldots$). In addition, if the inverted Majority is of interest, which is the case for $S$, $d_4$ must be $(n+1/2)\times \lambda$ and if the non-inverted output is required, which is the case for $C_o$, $d_5$ must be $n\lambda$. 
The AFA operation principle relies on a combined process of SWs propagation and interferences as follows: First, SWs are excited at $X$ and $Y$ and propagate diagonally until they interfere constructively or destructively depending on their phases at the connection point. Then, the resulting SW propagates and interferes constructively or destructively with the SW excited at $C_i$ at the next connection point. This interference result generates the final SW, which travels toward the outputs and  $\overline{MAJ(X,Y,C_i)}$ is detected at $S$ and $MAJ(X,Y,C_i)$ at $C_o$.
\begin{figure}[t]
\centering
  \includegraphics[width=0.9\linewidth]{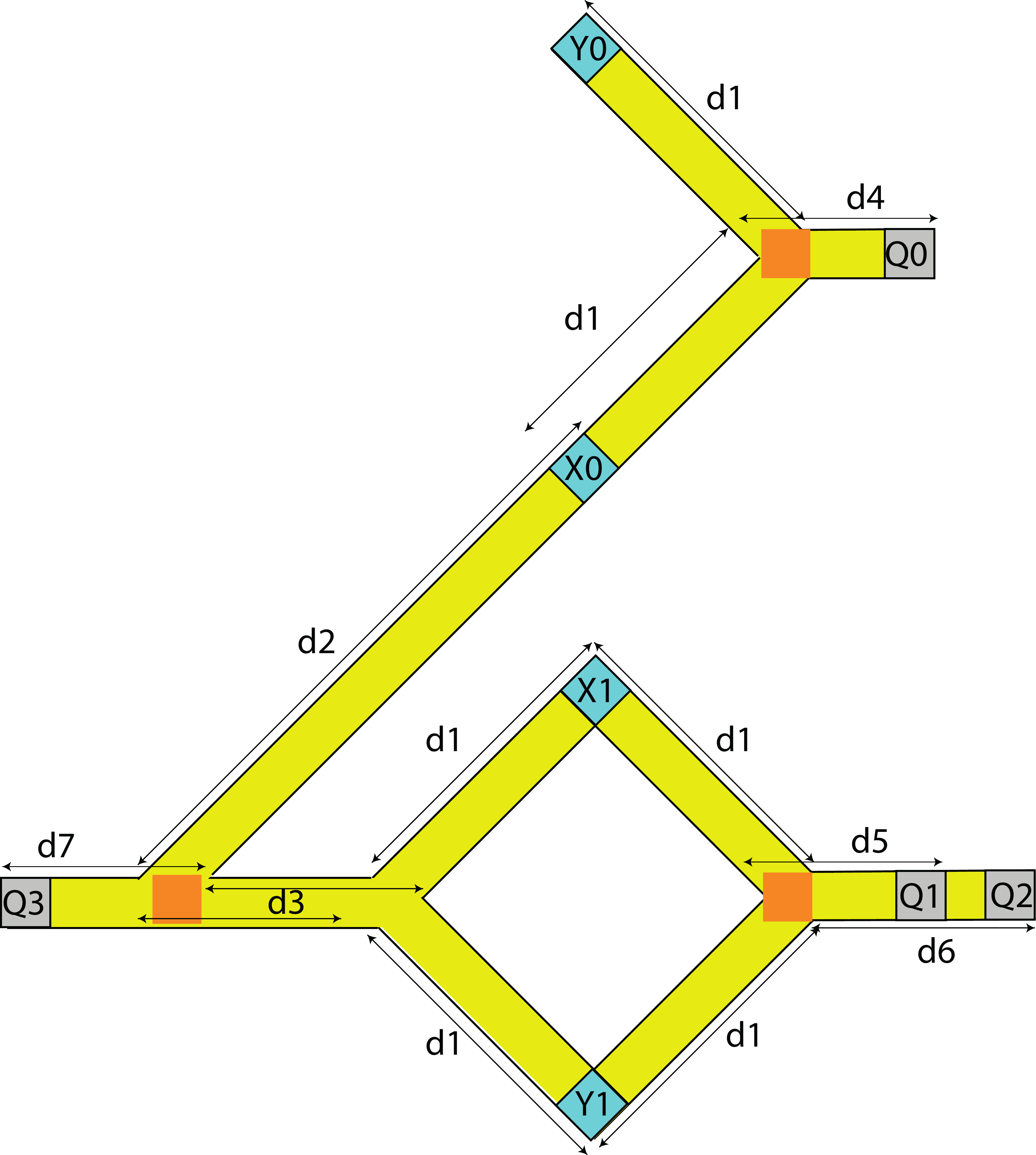}
  \caption{Approximate SW-based Multiplier}
  \label{fig:structure2}
\end{figure}

\begin{table}[t]
\caption{Accurate and Approximate SW-based Multiplier}
\label{table:2}
\centering
\scriptsize
  \begin{tabular}{|>{\centering}m{4.8em}|>{\centering}m{0.8em}|>{\centering}m{1.8em}|>{\centering}m{1.8em}|>{\centering}m{1.8em}|>{\centering}m{1.8em}|>{\centering}m{1.8em}|>{\centering}m{1.8em}|}
    \hline
   $X_1X_0Y_1Y_0$ & $Q_0$ & $Q_{1ac}$ & $Q_{1ap}$ & $Q_{2ac}$ & $Q_{2ap}$ & $Q_{3ac}$ & $Q_{3ap}$   \tabularnewline
    \hline
    $0$ $0$ $0$ $0$ & $0$ & $0$ & $0$ & $0$ & $0$ & $0$ & $0$  \tabularnewline
    \hline
    $0$  $0$  $0$  $1$ & $0$ & $0$ & $0$ & $0$ & $0$ & $0$ & $0$  \tabularnewline
    \hline
    $0$  $0$  $1$  $0$ & $0$ & $0$ & $0$ & $0$ & $0$ & $0$ & $0$  \tabularnewline
    \hline
    $0$  $0$  $1$  $1$ & $0$ & $0$ & $0$ & $0$ & $0$ & $0$ & $0$  \tabularnewline
    \hline
    $0$  $1$  $0$  $0$ & $0$ & $0$ & $0$ & $0$ & $0$ & $0$ & $0$  \tabularnewline
    \hline
    $0$  $1$ $0$  $1$ & $1$ & $0$ & $0$ & $0$ & $0$ & $0$ & $0$  \tabularnewline
    \hline
    $0$  $1$  $1$  $0$ & $0$ & $1$ & \textbf{\underline{0}} & $0$ & $0$ & $0$ & $0$  \tabularnewline
    \hline
    $0$  $1$  $1$  $1$ & $1$ & $1$ & \textbf{\underline{0}} & $0$ & $0$ & $0$ & $0$  \tabularnewline
    \hline
    $1$ $0$ $0$ $0$ & $0$ & $0$ & $0$ & $0$ & $0$ & $0$ & $0$  \tabularnewline
    \hline
    $1$ $0$ $0$ $1$ & $0$ & $1$ & \textbf{\underline{0}} & $0$ & $0$ & $0$ & $0$  \tabularnewline
    \hline
    $1$ $0$ $1$ $0$ & $0$ & $0$ & \textbf{\underline{1}} & $1$ & $1$ & $0$ & $0$  \tabularnewline
    \hline
    $1$ $0$ $1$ $1$ & $0$ & $1$ & $1$ & $1$ & $1$ & $0$ & $0$  \tabularnewline
    \hline
    $1$ $1$ $0$ $0$ & $0$ & $0$ & $0$ & $0$ & $0$ & $0$ & $0$  \tabularnewline
    \hline
    $1$ $1$ $0$ $1$ & $1$ & $1$ & \textbf{\underline{0}} & $0$ & $0$ & $0$ & $0$  \tabularnewline
    \hline
    $1$ $1$ $1$ $0$ & $0$ & $1$ & $1$ & $1$ & $1$ & $0$ & \textbf{\underline{1}}  \tabularnewline
    \hline
    $1$ $1$ $1$ $1$ & $1$ & $0$ & \textbf{\underline{1}} & $0$ & \textbf{\underline{1}} & $1$ & $1$  \tabularnewline
    \hline

    \end{tabular}
\end{table}

\subsection{SW Approximate $2$-bit inputs Multiplier}

Figure \ref{fig:structure2} presents the proposed Approximate $2$-bit inputs SW-based Multiplier (AMUL). Its inputs are the $2$-bit operands $X = (X_1, X_0)$ and $Y = (Y_1,Y_0)$ and its $4$-bit output is $Q=(Q_0, Q_1, Q_2, Q_3)$. AMUL consists of $3$ AND gates, which evaluate the AMUL outputs as  $Q_0=AND(X_0,Y_0)$, $Q_1=Q_2=AND(X_1,Y_1)$, and $Q_3=AND(X_0,X_1,Y_1)$, $4$ excitation cells, and $4$ detection cells. 

To evaluate the error rate we note that in the accurate MUL the outputs bits are computed as $Q_0=(X_0,Y_0)$, $Q_1=XOR(AND(X_0,Y_1),AND(X_1,Y_0))$, $Q_2=XOR(AND(AND(X_0,Y_1)$ $,AND(X_1,Y_0)),$ $AND(X_1,Y_1))$, and $Q_3=AND(AND(X_0,Y_0),$ $AND(X_1,Y_1))$, and present in Table \ref{table:2}  MUL and AMUL output values for all possible input combinations. Note that the erroneous values are written in bold and underlined. One can observe in the Table that AMUL computes $Q_0$ without any error, and $Q_1$, $Q_2$, and $Q_3$  with $31.25$\%,  $6.25$\%, and $6.25$\% error rate, respectively. 
However if threshold based output detection is utilized the error rate for $Q_1$ and $Q_3$ can be reduced to $25$\% and $0$\%, respectively,  as demonstrated in Section \ref{sec:Simulation Setup and Results}, which brings our proposal to an average error rate of $10$\%. 

The previously mentioned design parameters hold true for the AMUL as well. However, in contrast to AFA, AMUL relies on threshold based output detection, which means that the detection cells must be as close as possible to the last interference point, thus $d_4$, $d_5$, $d_6$, and $d_7$ should be minimized.

\section{Simulation Setup and Results}
\label{sec:Simulation Setup and Results}
The simulation setup and simulation results are provided and explained in this section.

\subsection{Simulation Setup}

We make use of a \SI{50}{nm} width and \SI{1}{nm} thick $Fe_{60}Co_{20}B_{20}$ waveguide and the parameters specified in Table \ref{table:3} \cite{parameters} to validate the proposed approximate designs (AFA and AMUL) by means of MuMax3 \cite{mumax}. As previously mentioned, the SW wavelength should be larger than the waveguide width to improve the interference pattern. Therefore, a \SI{55}{nm} SW wavelength was chosen. After that, the AFA dimension are determined as follows: $d_1$=\SI{330}{nm}, $d_2$=\SI{880}{nm}, $d_3$=\SI{220}{nm}, $d_4$=\SI{80}{nm}, and $d_5$=\SI{110}{nm} and the AMUL are $d_1$=\SI{330}{nm}, $d_2$=\SI{880}{nm}, $d_3$=\SI{220}{nm}, $d_4$=\SI{40}{nm}, $d_5$=\SI{40}{nm}, $d_6$=\SI{40}{nm}, and $d_7$=\SI{80}{nm}. Last, based on the SW dispersion relation, the SW frequency for a wavenumber $k$=$2\pi/\lambda$=\SI{50}{rad/\mu m} was calculated to correspond to a SW frequency of  \SI{10}{GHz}.

\begin{table}[t]
\caption{Simulation Parameters}
\label{table:3}
\centering
  \begin{tabular}{|c|c|}
    \hline
    Parameters & Values \\
    \hline
    Saturation magnetization $M_s$ & $1.1$ $\times$ $10^6$ A/m \\
    \hline
    Perpendicular anisotropy constant $k_{ani}$ & $0.83$ MJ/$m^3$\\
    \hline
    Damping constant $\alpha$ & $0.004$ \\
    \hline
    Exchange stiffness $A_{exch}$ & $18.5$ pJ/m \\
    \hline
  \end{tabular}
\end{table}

\begin{figure}[t]
\centering
  \includegraphics[width=\linewidth]{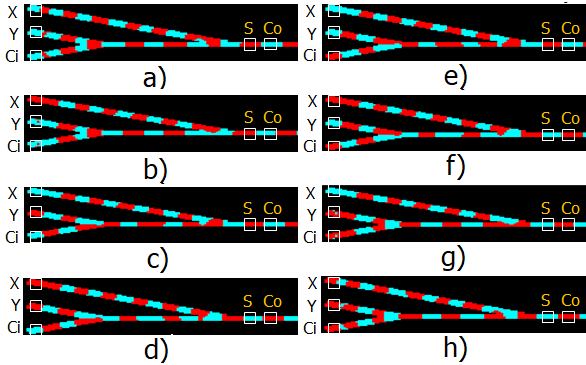}
  \caption{Approximate Spin Wave Based FA MuMax3 Simulation.}
  \label{fig:results1}
\end{figure}

\subsection{Simulation Results}

\subsubsection*{$1$-bit approximate FA based on phase detection} 

Figure \ref{fig:results1} a) to h) present AFA MuMax3 simulation results for \{$X$,$Y$,$C_i$\}= \{$0$,$0$,$0$\}, \{$0$,$0$,$0$\}, \{$0$,$0$,$1$\}, \{$0$,$1$,$0$\}, \{$0$,$1$,$1$\}, \{$1$,$0$,$0$\}, \{$1$,$0$,$1$\}, \{$1$,$1$,$0$\}, and \{$1$,$1$,$1$\}, respectively. Note that blue represents logic $0$ and red  logic $1$. One can observe in the Figure that the outputs $S$ and $C_{o}$ are detected as expected. For instance, $C_o=1$ for  \{$I_1$,$I_2$,$I_3$\}= \{$0$,$1$,$1$\}, \{$1$,$0$,$1$\}, \{$1$,$1$,$0$\}, and \{$1$,$1$,$1$\}, while $C_o=0$ for  \{$I_1$,$I_2$,$I_3$\}= \{$0$,$0$,$0$\}, \{$0$,$0$,$1$\}, \{$0$,$1$,$0$\}, and \{$1$,$0$,$0$\}. Moreover, $S$ is inverted $C_o=0$ as expected.

\begin{figure}[t]
\centering
  \includegraphics[width=\linewidth]{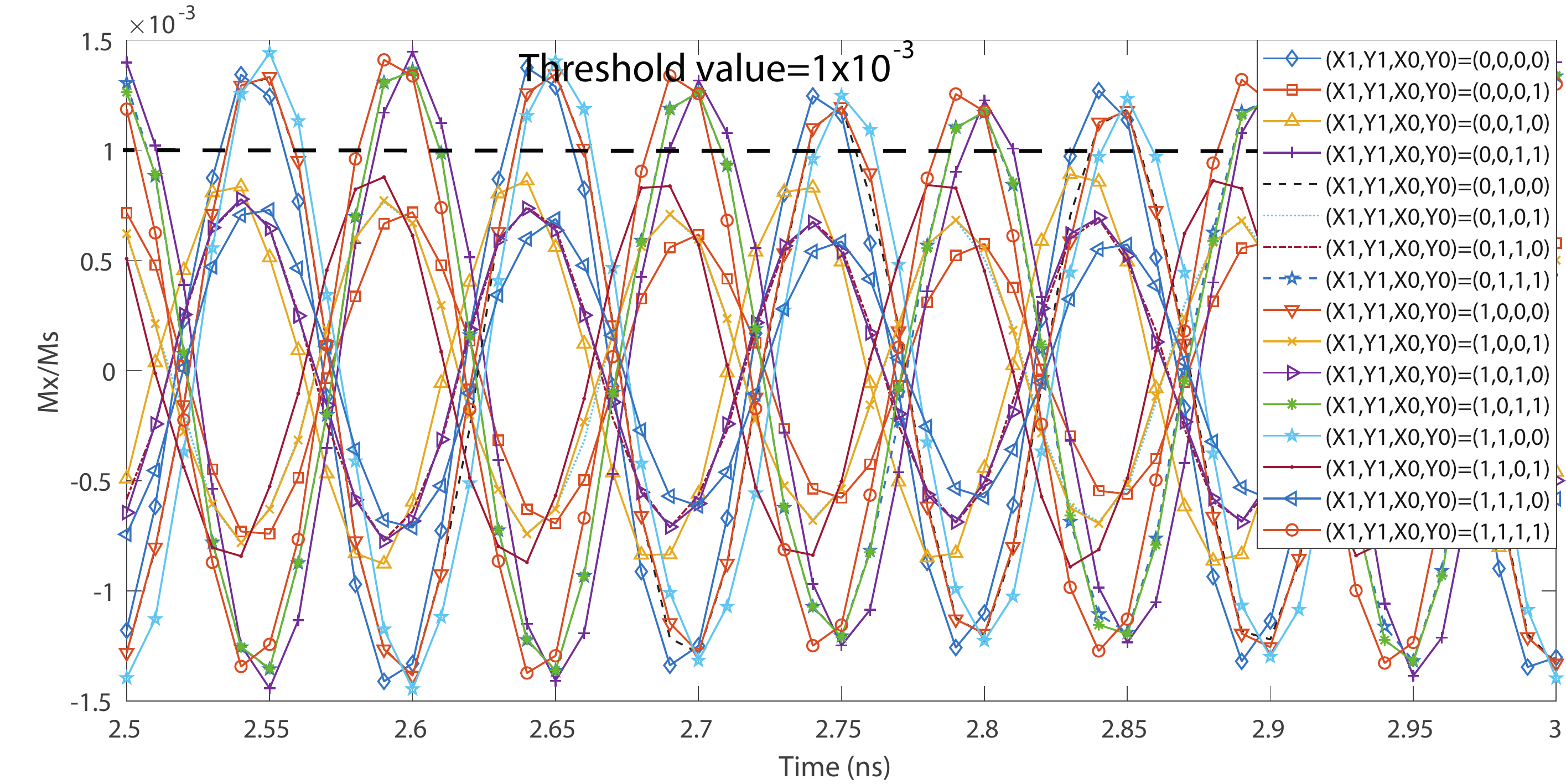}
  \caption{Normalized First AMUL Output.}
  \label{fig:results2}
\end{figure} 

\begin{figure}[t]
\centering
  \includegraphics[width=\linewidth]{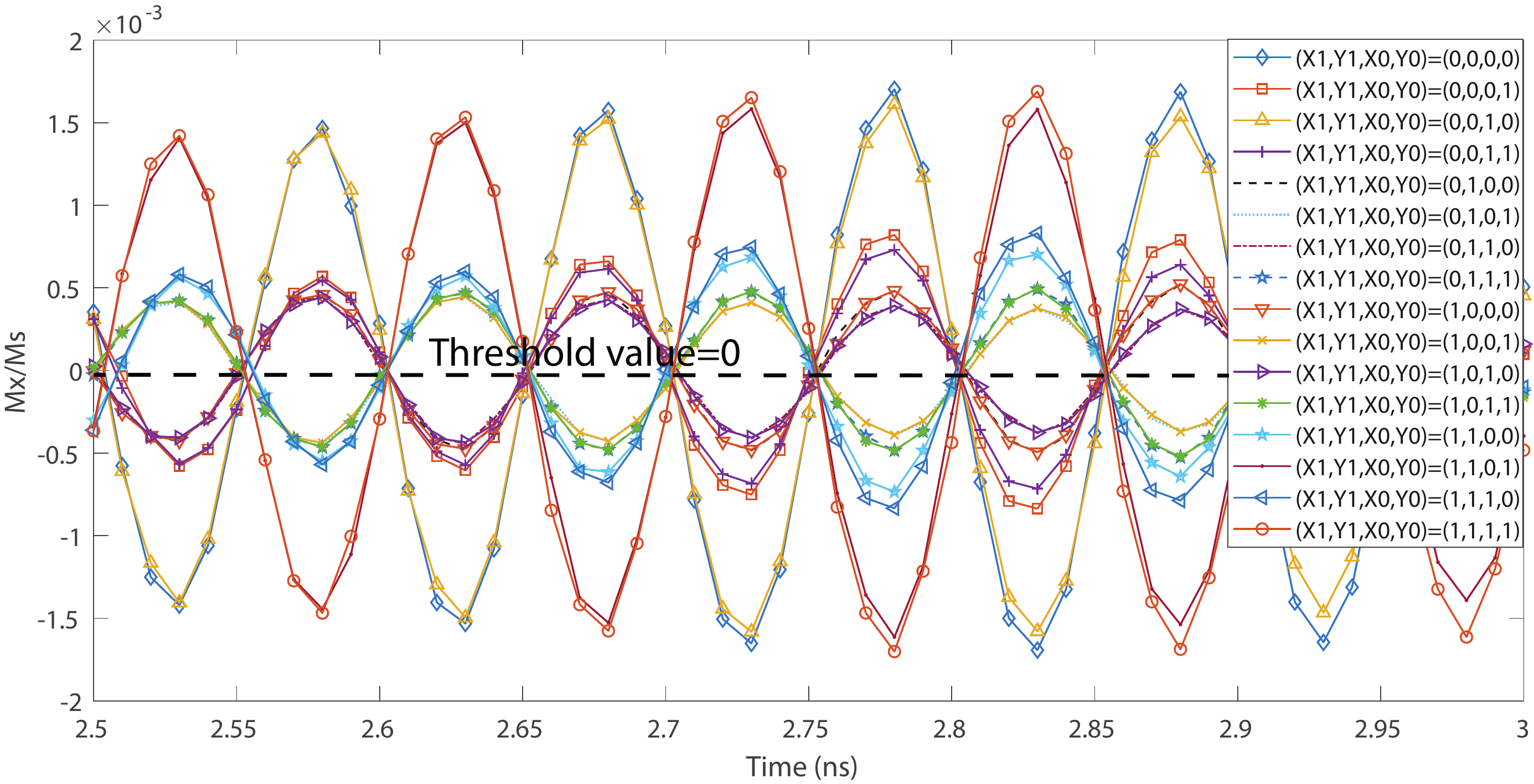}
  \caption{Normalized Second AMUL Output.}
  \label{fig:results3}
\end{figure} 

\begin{figure}[t]
\centering
  \includegraphics[width=\linewidth]{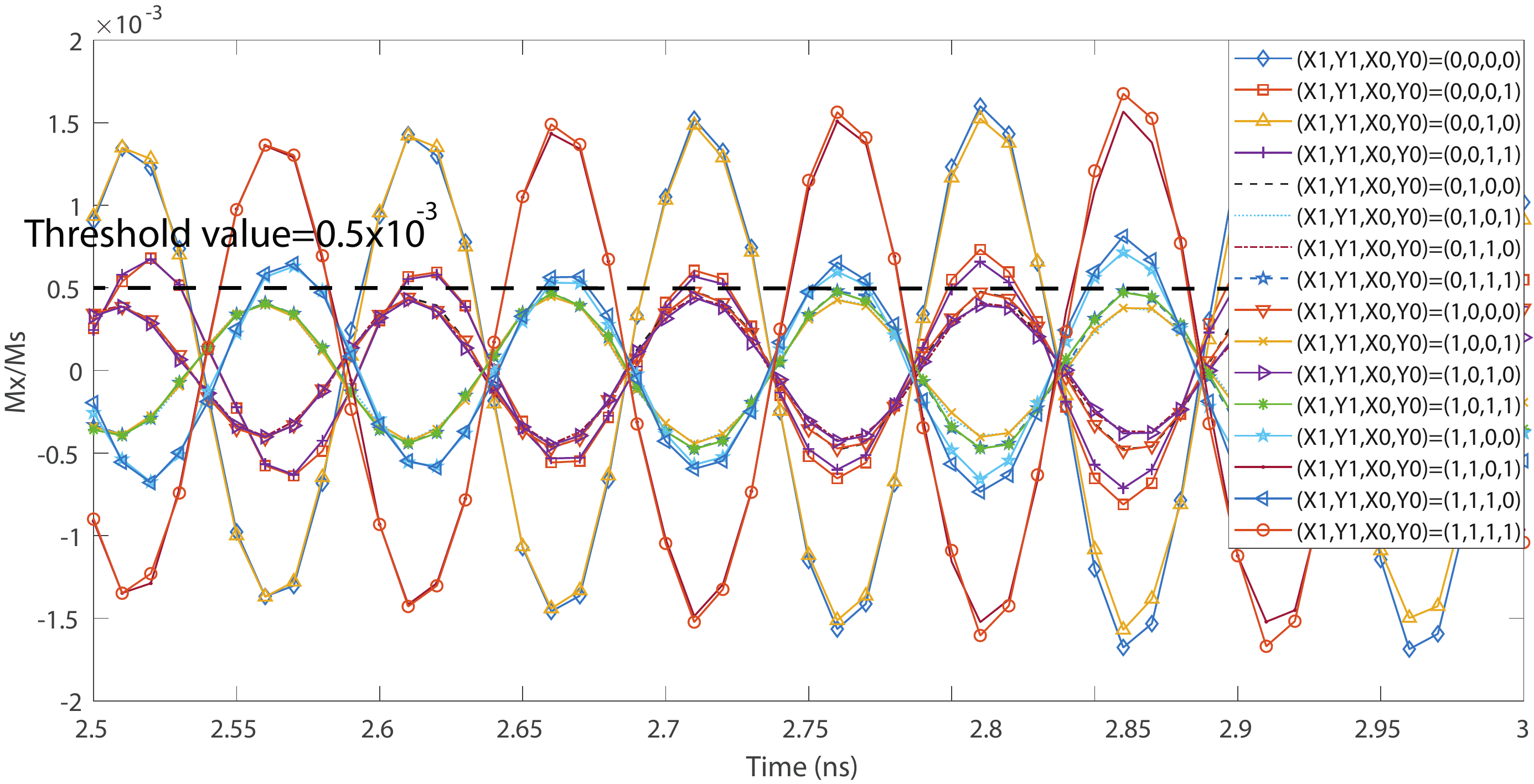}
  \caption{Normalized Third AMUL Output. }
  \label{fig:results4}
\end{figure} 

\begin{figure}[t]
\centering
  \includegraphics[width=\linewidth]{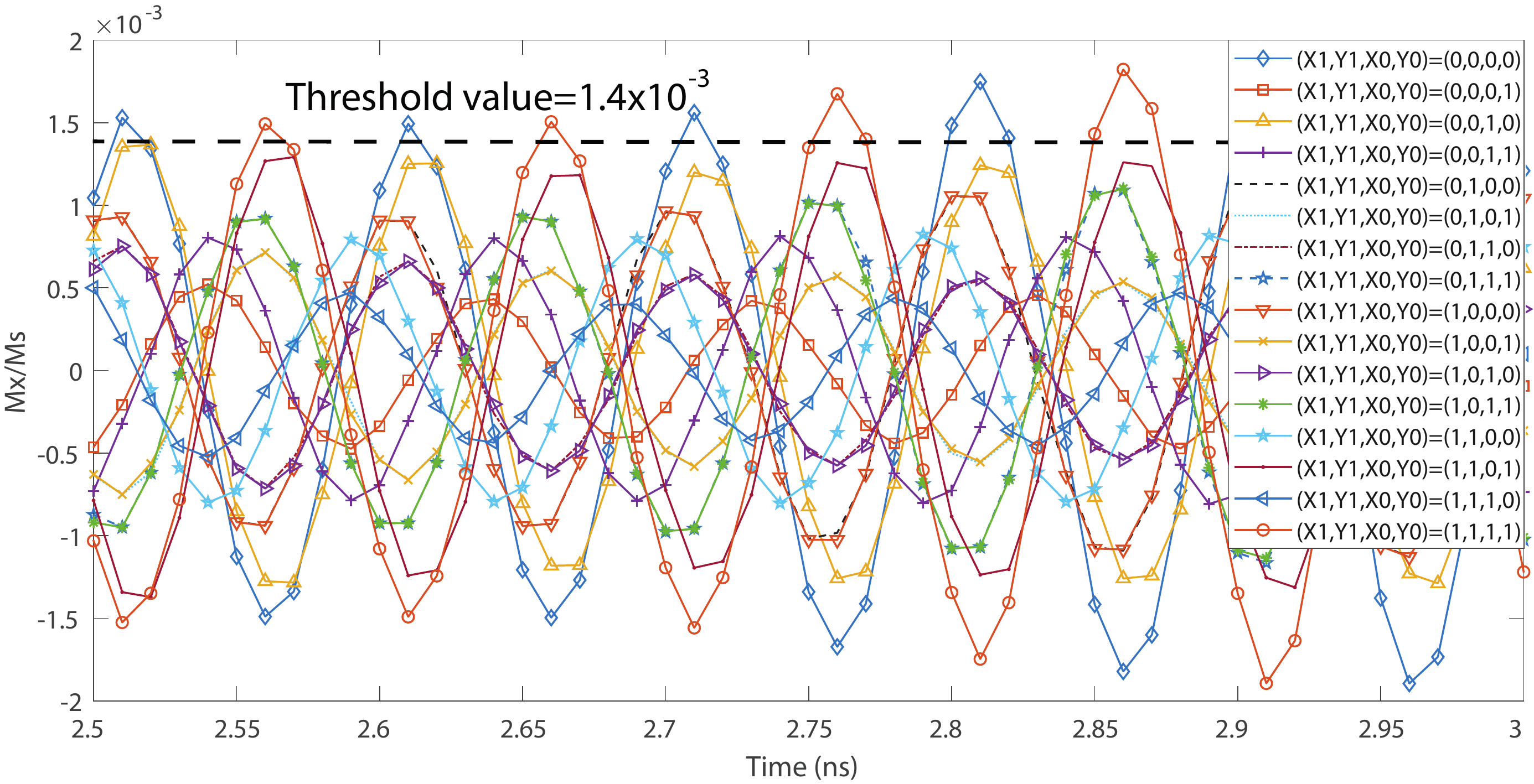}
  \caption{Normalized Forth AMUL Output. }
  \label{fig:results5}
\end{figure} 

\subsubsection*{$2$-bit inputs  approximate MUL based on threshold detection} 

Figures \ref{fig:results2} to \ref{fig:results5} present AMUL MuMax3 simulation results. In the figures, the $y$-axis presents the SWs $M_x$ over $M_s$ ratio, where $M_x$ is the magnetization projection along the $x$-direction and $M_s$ the saturation magnetization. Inspecting Figure \ref{fig:results2} we observe that $Q_1$ output SW magnetization at time \SI{2.7}{ns} for the input values $X_1Y_1X_0Y_0$=\{$0011$,$0111$,$1011$,$1111$\}, which should corresponds to  $Q_0 = 1$, is larger than $0.001M_s$ and smaller than $0.001M_s$ for the rest of the input combinations. Thus, by setting the detection threshold to $0.001M_s$, i.e., SW magnetization larger than $0.001M_s$ means logic $1$ and logic $0$ otherwise, $Q_0$ is always properly detected. 

Similarly, one can analyze Figure \ref{fig:results3}. For instance, the SWs magnetization for the input combinations $X_1Y_1X_0Y_0$=\{$0101$,$0111$,$1001$,$1011$,$1100$,$1101$,$1110$,$1111$\} are larger than $0$ when reading them at time \SI{2.76}{ns}, whereas for the other input combinations magnetization is less than $0$. Therefore, if the threshold is set to $0$ $Q_1$ value can be derived. 
Note that by doing so the theoretically predicted $Q_1$ error rate of $31.25$\% is diminished to $25$\%. 

Using the same way, Figure \ref{fig:results4} is analyzed. The SW magnetization for input combinations $X_1Y_1X_0Y_0$=\{$1100$,$1101$,$1110$,$1111$\} are larger than $0.0005M_s$ when reading them at time \SI{2.76}{ns}, whereas for the rest magnetization are less than $0.0005M_s$. Therefore, if the threshold is set to be $0.0005M_s$ $Q_2$ can be properly obtained with $0$\% error rate. 

Finally, Figure \ref{fig:results5} is analyzed in the same manner. The SWs magnetization for input combination $X_1Y_1X_0Y_0$=\{$1111$\} is larger than $0.0014M_s$ when reading them at time \SI{2.76}{ns}, whereas the rest of magnetization are less than $0.0014M_s$. Therefore, if the threshold is set to be $0.0014M_s$ $Q_3$ can be obtained with $0$\% error rate. 

\section{Performance Evaluation and Discussion}
\label{sec:Performance Evaluation}

In this section, the proposed AFA and AMUL are evaluated and compared with the state-of-the-art designs. Furthermore, the variability and thermal noise effects are discussed in addition to some open issues related to SW technology.

\subsection*{Performance Evaluation}

To get inside on the practical implications of our proposal we compare AFA with the state-of-the-art accurate SW \cite{SW2}, \SI{7}{nm} CMOS \cite{CMOS1}, SHE \cite{SHE2}, DWM \cite{DWM}, accurate and approximate \SI{45}{nm} CMOS \cite{approxCMOS1}, MTJ \cite{MTJ2}, and Spin-CMOS \cite{SPIN} counterparts in terms of energy, delay, and area (the number of utilized devices). To evaluate AFA we make use of the following assumptions: (i) Excitation and detection cells are Magnetoelectric (ME) cells which power consumption and delay are \SI{34}{nW} and \SI{0.42}{ns}, respectively \cite{Excitation_table_ref16}. (ii) During propagation and interference, SWs consume negligible amount of energy. (iii) The outputs are driving followup gates, the detection cells are not considered in the energy consumption calculation. (iv) Pulse signals are used to excite SWs. Note that due to SW  technology early stage development the aforementioned assumptions might need to be re-evaluated as the SW technology becomes more mature. 

The AFA delay is calculated by adding ME cell delay to the SW propagation delay through the waveguide determined by means of  micromagnetic simulation and equals to \SI{1.84}{ns}. Table \ref{table:4} presents the results of the evaluation and comparison. Inspecting the Table, it is clear that AFA outperforms state-of-the-art \SI{7}{nm} CMOS \cite{CMOS1} accurate FA by energy reductions of approximately $33$\%, while exhibiting more than $2$ orders of magnitude larger delay. Furthermore, AFA saves approximately $69$\% and $44$\% energy while requiring $15$x and $18$x larger delay when compared with \SI{45}{nm} CMOS based accurate and approximate FA, respectively, while having the same error rate as the approximate FA in \cite{approxCMOS1}. When compared with other emerging technologies based designs, AFA consumes $5$ orders of magnitude less energy than MTJ based accurate and approximate FAs while exhibiting $42$\% lower delay and having $50$\% better error rate than the MTJ approximate FA in \cite{MTJ2}. Moreover, AFA consumes $5$ and $3$ orders of magnitude less energy than SHE- and DWM- based accurate FAs, respectively, has $3.8$x lower and $52$\% more delay than SHE  \cite{SHE2}  and  DWM  \cite{DWM} based FAs, respectively. Furthermore, AFA consumes approximately $4$ and $3$ orders of magnitude less energy while requiring $38$\% and $8$\% lower delay in comparison with the accurate and approximate Spin-CMOS based FAs, respectively, while having the same error rate as the approximate FA in \cite{SPIN}. Last but not least, AFA outperforms the SW based accurate FA \cite{SW2} by $40$\% and $35$\% in terms of energy and delay, respectively. Note that as a chip real-estate estimation that the proposed approximate FA requires the lowest number of devices.

\begin{table}[t]
\caption{Full Adder Performance Comparison}
\label{table:4}
\centering
  \begin{tabular}{|c|c|c|c|c|c|}
    \hline
     Technology & Type & Error Rate & Energy (fJ)  &  Delay (ns) & Device No. \tabularnewline
    \hline
    CMOS \cite{CMOS1}& Accurate  & $0$  & $0.065$ & $0.005$ & $28$ \tabularnewline \hline
     CMOS \cite{approxCMOS1}& Accurate  & $0$  & $0.14$ & $0.12$ & $24$ \tabularnewline \hline
     CMOS \cite{approxCMOS1}& Approximate  & $0.25$ & $0.077$  & $0.1$ & $14$ \tabularnewline \hline
     MTJ \cite{MTJ2}& Accurate  & $0$ & $5685$  & $3.019$ & $29$ \tabularnewline \hline
     MTJ \cite{MTJ2}& Approximate  & $0.5$ & $5109$  & $3.016$ & $25$ \tabularnewline \hline
     MTJ \cite{MTJ2}& Approximate  & $0.5$ & $2471$  & $3.152$ & $29$ \tabularnewline \hline
     SHE \cite{SHE2}& Accurate  & $0$ & $4970$  & $7$ & $26$ \tabularnewline \hline
     DWM \cite{DWM}& Accurate  & $0$ & $74.5$  & $0.877$ & $26$ \tabularnewline \hline
     Spin CMOS \cite{SPIN}& Accurate  & $0$ & $166.7$  & $3$ & $34$ \tabularnewline \hline
     Spin CMOS \cite{SPIN}& Approximate  & $0.25$ & $58$  & $2$ & $34$ \tabularnewline \hline
     Spin Wave \cite{SW2} & Accurate  & $0$ & $0.072$  & $2.86$ & $7$ \tabularnewline \hline
     Spin Wave & Approximate  & $0.25$ & $0.043$  & $1.84$ & $5$ \tabularnewline \hline

  \end{tabular}
\end{table}

Under the same assumptions AMUL delay is \SI{3.3}{ns} and we compare it  with state-of-the-art SW \cite{amahmoud1} and CMOS \cite{approxCMOS2} counterparts.
As delay figures are not mentioned for the approximate multiplier in \cite{approxCMOS2}, its energy consumption was estimated based on the \SI{16}{nm}CMOS figures provided in \cite{16nmCMOS}.
Table \ref{table:5} present the results of the evaluation and comparison. Inspecting the Table, it is clear that AMUL outperforms accurate \SI{16}{nm} CMOS \cite{approxCMOS2} and approximate \SI{16}{nm} CMOS \cite{approxCMOS2} counterparts by diminishing the energy consumption by $16$x and $5$x while exhibiting $33$x and $55$x larger delay, respectively. AMUL has an average error rate of $10$\% while $12.5$\% is the average error rate for the approximate CMOS  counterpart \cite{approxCMOS2}. Note that the average error rate is calculated by adding the average of $Q_1$, $Q_2$, and $Q_3$ error rates  as the first output $Q_0$ is accurately compute in both implementations.  
When compared with accurate MUL SW implementations, AMUL  saves $4$x and $6$x energy and approximately $6$x lower and $2$x more delay in comparison with SW coupler and conversion based MUL implementations, respectively.  We note that the SW propagation delay is neglected into the evaluation of the SW conversion based MUL in \cite{amahmoud1}. One can observe from the Table that the proposed MUL requires less ME cells than the SW designs in \cite{amahmoud1} which indicates  that the design in \cite{amahmoud1} has a larger area and by implication a larger delay if SW propagation is also considered. 

\begin{table}[t]
\caption{$2$-bit inputs Multiplier Performance Comparison}
\label{table:5}
\scriptsize{
\centering
  \begin{tabular}{|c|c|c|c|c|c|}
    \hline
     Design & \multicolumn{2}{c|}{CMOS\cite{16nmCMOS,approxCMOS2}} & \multicolumn{2}{c|}{SW\cite{amahmoud1}} &  Proposed SW MUL  \tabularnewline
    \hline
     Implemented method & \multicolumn{2}{c|}{-} & Coupler Cascading & Conversion Cascading & -    \tabularnewline \hline
     Type  & Accurate & Approximate & Accurate & Accurate & Approximate  \tabularnewline \hline
     Average Error Rate & $0$ & $0.125$ & $0$ & $0$ & $0.1$  \tabularnewline \hline
     energy (aJ) & $959$ & $300$ & $259$ & $374$ & $58$ \tabularnewline \hline
     Delay (ns) & $0.1$ & $0.06$ & $21$ & $1.68$ & $3.3$ \tabularnewline \hline
     Device No. & $52$ & $30$ & $22$ & $30$ & $8$ \tabularnewline \hline
  \end{tabular}}
\end{table}

\subsection*{Variability and Thermal Effect}
In this paper, the main target is to propose and validate by means of micromagnetic simulations the approximate FA and MUL as proof of the concepts without considering the impacts of the thermal noise and the variability. However, it was reported that the thermal noise has limited effect on the gate function and consequently the gate works correctly at different temperature \cite{DC}. In addition, the effect of the edge roughness and the waveguide trapezoidal cross section were demonstrated \cite{DC}. It was suggested that both effects are very small and the gate operates correctly at their presence as well \cite{DC}. Therefore, we don't expect neither the thermal noise nor the geometrical variability to have large impact on the proposed circuits. However, we plan to investigate these phenomena in the future.

\subsection*{Discussion}
Although the evaluation demonstrated that the SW technology has the needed requirements to improve the state-of-the-art in terms of energy as well as area consumption, but a number of open issues are still to be solved \cite{amahmoud2}:

\begin{itemize}
 \item 
 Immature technology: It seems that the ME cells are the right option to excite and detect the SW because of their ultra low energy consumption, acceptable delay and scalability. However, ME cells are not realized experimentally until now. 
 \item 
Scalability: In terms of area SW circuit have a great scaling potential as for proper functionality SW device dimensions must be greater or equal than the SW wavelength, which can reach down to the $nm$ range. Several SW circuit area benchmarkings have been reported \cite{Excitation_table_ref16} which indicate that hybrid spin-wave--CMOS circuits have very small area. Although the assumptions the benchmarking is based on might not be fully realistic, they give an indication regarding the expected area. For example, the area of a $32$-bit divider (DIV32) implemented in hybrid SW-CMOS is roughly about $3.5$x smaller than the one of the 10 nm CMOS counterpart. However, few things are needed before being able to realize nano-scale SW device such as excitation and detection: currently, it is not possible to distinguish $nm$ SWs from noise.
\end{itemize}

\section{Conclusions}
\label{sec:Conclusion}

We proposed and validated by means of micromagnetic simulations a novel approximate energy efficient spin wave based Full Adder (AFA) and $2$-bit inputs multiplier (AMUL). Both designs were evaluated and compared with the state-of-the-art counterparts. AFA saves $43$\% and $33$\% energy when compared with the state-of-the-art SW and \SI{7}{nm} CMOS, respectively, and $69$\% and $44$\% in comparison with accurate and approximate \SI{45}{nm} CMOS, respectively. In addition, it saves more than $2$ orders of magnitude when compared with accurate SHE, and accurate and approximate DWM, MTJ, and Spin-CMOS FAs. Moreover, it achieves the same error rate as approximate \SI{45}{nm} CMOS and Spin-CMOS FA whereas it exhibits $50$\% less error rate than approximate DWM FA and requires at least $29$\% less chip real-estate in comparison with the other state-of-the-art designs. 
At its turn AMUL saves at least $2$x and $5$x energy in comparison with the state-of-the-art accurate SW designs and \SI{16}{nm} CMOS accurate and approximate designs, respectively. Moreover, the AMUL has an average error rate of $10$\%, while the approximate CMOS MUL has an average error rate of $12.5$\%, and requires at least $64$\% less chip real-estate.

\section*{Acknowledgement}
This work has received funding from the European Union's Horizon 2020 research and innovation program within the FET-OPEN project CHIRON under grant agreement No. 801055. It has also been partially supported by imec's industrial affiliate program on beyond-CMOS logic. F.V. acknowledges financial support from Flanders Research Foundation (FWO) through grant No.~1S05719N.

\bibliography{Spin_Wave_Based_Approximate_Computing}

\begin{thebibliography}{45}%
\makeatletter
\providecommand \@ifxundefined [1]{%
 \@ifx{#1\undefined}
}%
\providecommand \@ifnum [1]{%
 \ifnum #1\expandafter \@firstoftwo
 \else \expandafter \@secondoftwo
 \fi
}%
\providecommand \@ifx [1]{%
 \ifx #1\expandafter \@firstoftwo
 \else \expandafter \@secondoftwo
 \fi
}%
\providecommand \natexlab [1]{#1}%
\providecommand \enquote  [1]{``#1''}%
\providecommand \bibnamefont  [1]{#1}%
\providecommand \bibfnamefont [1]{#1}%
\providecommand \citenamefont [1]{#1}%
\providecommand \href@noop [0]{\@secondoftwo}%
\providecommand \href [0]{\begingroup \@sanitize@url \@href}%
\providecommand \@href[1]{\@@startlink{#1}\@@href}%
\providecommand \@@href[1]{\endgroup#1\@@endlink}%
\providecommand \@sanitize@url [0]{\catcode `\\12\catcode `\$12\catcode
  `\&12\catcode `\#12\catcode `\^12\catcode `\_12\catcode `\%12\relax}%
\providecommand \@@startlink[1]{}%
\providecommand \@@endlink[0]{}%
\providecommand \url  [0]{\begingroup\@sanitize@url \@url }%
\providecommand \@url [1]{\endgroup\@href {#1}{\urlprefix }}%
\providecommand \urlprefix  [0]{URL }%
\providecommand \Eprint [0]{\href }%
\providecommand \doibase [0]{http://dx.doi.org/}%
\providecommand \selectlanguage [0]{\@gobble}%
\providecommand \bibinfo  [0]{\@secondoftwo}%
\providecommand \bibfield  [0]{\@secondoftwo}%
\providecommand \translation [1]{[#1]}%
\providecommand \BibitemOpen [0]{}%
\providecommand \bibitemStop [0]{}%
\providecommand \bibitemNoStop [0]{.\EOS\space}%
\providecommand \EOS [0]{\spacefactor3000\relax}%
\providecommand \BibitemShut  [1]{\csname bibitem#1\endcsname}%
\let\auto@bib@innerbib\@empty
\bibitem [{\citenamefont {Shah}, \citenamefont {Steyerberg},\ and\
  \citenamefont {Kent}(2018)}]{data1}%
  \BibitemOpen
  \bibfield  {author} {\bibinfo {author} {\bibfnamefont {N.~D.}\ \bibnamefont
  {Shah}}, \bibinfo {author} {\bibfnamefont {E.~W.}\ \bibnamefont
  {Steyerberg}}, \ and\ \bibinfo {author} {\bibfnamefont {D.~M.}\ \bibnamefont
  {Kent}},\ }\href@noop {} {\bibfield  {journal} {\bibinfo  {journal} {JAMA}\ }
  (\bibinfo {year} {2018})}\BibitemShut {NoStop}%
\bibitem [{\citenamefont {Haron}\ and\ \citenamefont
  {Hamdioui}(2008)}]{cmosscaling1}%
  \BibitemOpen
  \bibfield  {author} {\bibinfo {author} {\bibfnamefont {N.~Z.}\ \bibnamefont
  {Haron}}\ and\ \bibinfo {author} {\bibfnamefont {S.}~\bibnamefont
  {Hamdioui}},\ }in\ \href@noop {} {\emph {\bibinfo {booktitle} {Design and
  Test Workshop, 2008. IDT 2008. 3rd International}}}\ (\bibinfo {organization}
  {IEEE},\ \bibinfo {year} {2008})\ pp.\ \bibinfo {pages} {98--103}\BibitemShut
  {NoStop}%
\bibitem [{\citenamefont {Yu}\ \emph {et~al.}(2020)\citenamefont {Yu},
  \citenamefont {Nane}, \citenamefont {Ashraf}, \citenamefont {Taouil},
  \citenamefont {Hamdioui}, \citenamefont {Corporaal},\ and\ \citenamefont
  {Bertels}}]{mem1}%
  \BibitemOpen
  \bibfield  {author} {\bibinfo {author} {\bibfnamefont {J.}~\bibnamefont
  {Yu}}, \bibinfo {author} {\bibfnamefont {R.}~\bibnamefont {Nane}}, \bibinfo
  {author} {\bibfnamefont {I.}~\bibnamefont {Ashraf}}, \bibinfo {author}
  {\bibfnamefont {M.}~\bibnamefont {Taouil}}, \bibinfo {author} {\bibfnamefont
  {S.}~\bibnamefont {Hamdioui}}, \bibinfo {author} {\bibfnamefont
  {H.}~\bibnamefont {Corporaal}}, \ and\ \bibinfo {author} {\bibfnamefont
  {K.}~\bibnamefont {Bertels}},\ }\href {\doibase 10.1109/TETC.2017.2760927}
  {\bibfield  {journal} {\bibinfo  {journal} {IEEE Transactions on Emerging
  Topics in Computing}\ }\textbf {\bibinfo {volume} {8}},\ \bibinfo {pages}
  {545} (\bibinfo {year} {2020})}\BibitemShut {NoStop}%
\bibitem [{\citenamefont {Vourkas}, \citenamefont {Stathis},\ and\
  \citenamefont {Sirakoulis}(2018)}]{mem2}%
  \BibitemOpen
  \bibfield  {author} {\bibinfo {author} {\bibfnamefont {I.}~\bibnamefont
  {Vourkas}}, \bibinfo {author} {\bibfnamefont {D.}~\bibnamefont {Stathis}}, \
  and\ \bibinfo {author} {\bibfnamefont {G.~H.}\ \bibnamefont {Sirakoulis}},\
  }\href {\doibase 10.1109/TETC.2015.2420353} {\bibfield  {journal} {\bibinfo
  {journal} {IEEE Transactions on Emerging Topics in Computing}\ }\textbf
  {\bibinfo {volume} {6}},\ \bibinfo {pages} {145} (\bibinfo {year}
  {2018})}\BibitemShut {NoStop}%
\bibitem [{\citenamefont {Maestro-Izquierdo}\ \emph {et~al.}(2019)\citenamefont
  {Maestro-Izquierdo}, \citenamefont {Martin-Martinez}, \citenamefont {Yepes},
  \citenamefont {Escudero}, \citenamefont {Rodriguez}, \citenamefont {Nafria},
  \citenamefont {Aymerich},\ and\ \citenamefont {Rubio}}]{mem3}%
  \BibitemOpen
  \bibfield  {author} {\bibinfo {author} {\bibfnamefont {M.}~\bibnamefont
  {Maestro-Izquierdo}}, \bibinfo {author} {\bibfnamefont {J.}~\bibnamefont
  {Martin-Martinez}}, \bibinfo {author} {\bibfnamefont {A.~C.}\ \bibnamefont
  {Yepes}}, \bibinfo {author} {\bibfnamefont {M.}~\bibnamefont {Escudero}},
  \bibinfo {author} {\bibfnamefont {R.}~\bibnamefont {Rodriguez}}, \bibinfo
  {author} {\bibfnamefont {M.}~\bibnamefont {Nafria}}, \bibinfo {author}
  {\bibfnamefont {X.}~\bibnamefont {Aymerich}}, \ and\ \bibinfo {author}
  {\bibfnamefont {A.}~\bibnamefont {Rubio}},\ }\href {\doibase
  10.1109/TETC.2017.2760929} {\bibfield  {journal} {\bibinfo  {journal} {IEEE
  Transactions on Emerging Topics in Computing}\ }\textbf {\bibinfo {volume}
  {7}},\ \bibinfo {pages} {545} (\bibinfo {year} {2019})}\BibitemShut {NoStop}%
\bibitem [{\citenamefont {Pouyan}, \citenamefont {Amat},\ and\ \citenamefont
  {Rubio}(2018)}]{mem4}%
  \BibitemOpen
  \bibfield  {author} {\bibinfo {author} {\bibfnamefont {P.}~\bibnamefont
  {Pouyan}}, \bibinfo {author} {\bibfnamefont {E.}~\bibnamefont {Amat}}, \ and\
  \bibinfo {author} {\bibfnamefont {A.}~\bibnamefont {Rubio}},\ }\href
  {\doibase 10.1109/TETC.2016.2581700} {\bibfield  {journal} {\bibinfo
  {journal} {IEEE Transactions on Emerging Topics in Computing}\ }\textbf
  {\bibinfo {volume} {6}},\ \bibinfo {pages} {207} (\bibinfo {year}
  {2018})}\BibitemShut {NoStop}%
\bibitem [{\citenamefont {{Jiang}}\ \emph {et~al.}(2019)\citenamefont
  {{Jiang}}, \citenamefont {{Laurenciu}}, \citenamefont {{Wang}},\ and\
  \citenamefont {{Cotofana}}}]{Yande1}%
  \BibitemOpen
  \bibfield  {author} {\bibinfo {author} {\bibfnamefont {Y.}~\bibnamefont
  {{Jiang}}}, \bibinfo {author} {\bibfnamefont {N.~C.}\ \bibnamefont
  {{Laurenciu}}}, \bibinfo {author} {\bibfnamefont {H.}~\bibnamefont {{Wang}}},
  \ and\ \bibinfo {author} {\bibfnamefont {S.~D.}\ \bibnamefont {{Cotofana}}},\
  }\href@noop {} {\bibfield  {journal} {\bibinfo  {journal} {IEEE Transactions
  on Nanotechnology}\ }\textbf {\bibinfo {volume} {18}},\ \bibinfo {pages}
  {287} (\bibinfo {year} {2019})}\BibitemShut {NoStop}%
\bibitem [{\citenamefont {Banadaki}\ and\ \citenamefont
  {Srivastava}(2015)}]{gra1}%
  \BibitemOpen
  \bibfield  {author} {\bibinfo {author} {\bibfnamefont {Y.}~\bibnamefont
  {Banadaki}}\ and\ \bibinfo {author} {\bibfnamefont {A.}~\bibnamefont
  {Srivastava}},\ }\href {\doibase 10.1109/TETC.2015.2445104} {\bibfield
  {journal} {\bibinfo  {journal} {IEEE Transactions on Emerging Topics in
  Computing}\ }\textbf {\bibinfo {volume} {3}},\ \bibinfo {pages} {458}
  (\bibinfo {year} {2015})}\BibitemShut {NoStop}%
\bibitem [{\citenamefont {Nishad}\ and\ \citenamefont {Sharma}(2015)}]{gra2}%
  \BibitemOpen
  \bibfield  {author} {\bibinfo {author} {\bibfnamefont {A.}~\bibnamefont
  {Nishad}}\ and\ \bibinfo {author} {\bibfnamefont {R.}~\bibnamefont
  {Sharma}},\ }\href {\doibase 10.1109/TETC.2015.2486748} {\bibfield  {journal}
  {\bibinfo  {journal} {IEEE Transactions on Emerging Topics in Computing}\
  }\textbf {\bibinfo {volume} {3}},\ \bibinfo {pages} {470} (\bibinfo {year}
  {2015})}\BibitemShut {NoStop}%
\bibitem [{\citenamefont {Agarwal}\ \emph {et~al.}(2018)\citenamefont
  {Agarwal}, \citenamefont {Burr}, \citenamefont {Chen}, \citenamefont {Das},
  \citenamefont {Debenedictis}, \citenamefont {Frank}, \citenamefont {Franzon},
  \citenamefont {Holmes}, \citenamefont {Marinella},\ and\ \citenamefont
  {Rakshit}}]{ITRS}%
  \BibitemOpen
  \bibfield  {author} {\bibinfo {author} {\bibfnamefont {S.}~\bibnamefont
  {Agarwal}}, \bibinfo {author} {\bibfnamefont {G.}~\bibnamefont {Burr}},
  \bibinfo {author} {\bibfnamefont {A.}~\bibnamefont {Chen}}, \bibinfo {author}
  {\bibfnamefont {S.}~\bibnamefont {Das}}, \bibinfo {author} {\bibfnamefont
  {E.}~\bibnamefont {Debenedictis}}, \bibinfo {author} {\bibfnamefont {M.~P.}\
  \bibnamefont {Frank}}, \bibinfo {author} {\bibfnamefont {P.}~\bibnamefont
  {Franzon}}, \bibinfo {author} {\bibfnamefont {S.}~\bibnamefont {Holmes}},
  \bibinfo {author} {\bibfnamefont {M.}~\bibnamefont {Marinella}}, \ and\
  \bibinfo {author} {\bibfnamefont {T.}~\bibnamefont {Rakshit}},\ }\href@noop
  {} {\enquote {\bibinfo {title} {International roadmap of devices and systems
  2017 edition: Beyond cmos chapter.}}\ }\bibinfo {type} {Tech. Rep.}\
  (\bibinfo  {institution} {Sandia National Lab.(SNL-NM), Albuquerque, NM
  (United States)},\ \bibinfo {year} {2018})\BibitemShut {NoStop}%
\bibitem [{\citenamefont {Zabihi}\ \emph {et~al.}(2019)\citenamefont {Zabihi},
  \citenamefont {Chowdhury}, \citenamefont {Zhao}, \citenamefont {Karpuzcu},
  \citenamefont {Wang},\ and\ \citenamefont {Sapatnekar}}]{spin1}%
  \BibitemOpen
  \bibfield  {author} {\bibinfo {author} {\bibfnamefont {M.}~\bibnamefont
  {Zabihi}}, \bibinfo {author} {\bibfnamefont {Z.}~\bibnamefont {Chowdhury}},
  \bibinfo {author} {\bibfnamefont {Z.}~\bibnamefont {Zhao}}, \bibinfo {author}
  {\bibfnamefont {U.~R.}\ \bibnamefont {Karpuzcu}}, \bibinfo {author}
  {\bibfnamefont {J.}~\bibnamefont {Wang}}, \ and\ \bibinfo {author}
  {\bibfnamefont {S.~S.}\ \bibnamefont {Sapatnekar}},\ }\href {\doibase
  10.1109/TC.2018.2858251} {\bibfield  {journal} {\bibinfo  {journal} {IEEE
  Transactions on Computers}\ }\textbf {\bibinfo {volume} {68}},\ \bibinfo
  {pages} {1159} (\bibinfo {year} {2019})}\BibitemShut {NoStop}%
\bibitem [{\citenamefont {Bai}\ \emph {et~al.}(2018)\citenamefont {Bai},
  \citenamefont {DeMara}, \citenamefont {Di},\ and\ \citenamefont
  {Lin}}]{spin2}%
  \BibitemOpen
  \bibfield  {author} {\bibinfo {author} {\bibfnamefont {Y.}~\bibnamefont
  {Bai}}, \bibinfo {author} {\bibfnamefont {R.~F.}\ \bibnamefont {DeMara}},
  \bibinfo {author} {\bibfnamefont {J.}~\bibnamefont {Di}}, \ and\ \bibinfo
  {author} {\bibfnamefont {M.}~\bibnamefont {Lin}},\ }\href {\doibase
  10.1109/TC.2017.2776139} {\bibfield  {journal} {\bibinfo  {journal} {IEEE
  Transactions on Computers}\ }\textbf {\bibinfo {volume} {67}},\ \bibinfo
  {pages} {631} (\bibinfo {year} {2018})}\BibitemShut {NoStop}%
\bibitem [{\citenamefont {Vyas}\ \emph {et~al.}(2021)\citenamefont {Vyas},
  \citenamefont {Jiang-Wei}, \citenamefont {Zhou}, \citenamefont {Hu},\ and\
  \citenamefont {Friedman}}]{spin3}%
  \BibitemOpen
  \bibfield  {author} {\bibinfo {author} {\bibfnamefont {V.}~\bibnamefont
  {Vyas}}, \bibinfo {author} {\bibfnamefont {L.}~\bibnamefont {Jiang-Wei}},
  \bibinfo {author} {\bibfnamefont {P.}~\bibnamefont {Zhou}}, \bibinfo {author}
  {\bibfnamefont {X.}~\bibnamefont {Hu}}, \ and\ \bibinfo {author}
  {\bibfnamefont {J.~S.}\ \bibnamefont {Friedman}},\ }\href {\doibase
  10.1109/TC.2020.2986970} {\bibfield  {journal} {\bibinfo  {journal} {IEEE
  Transactions on Computers}\ }\textbf {\bibinfo {volume} {70}},\ \bibinfo
  {pages} {128} (\bibinfo {year} {2021})}\BibitemShut {NoStop}%
\bibitem [{\citenamefont {Mahmoud}\ \emph {et~al.}(2020)\citenamefont
  {Mahmoud}, \citenamefont {Ciubotaru}, \citenamefont {Vanderveken},
  \citenamefont {Chumak}, \citenamefont {Hamdioui}, \citenamefont {Adelmann},\
  and\ \citenamefont {Cotofana}}]{amahmoud2}%
  \BibitemOpen
  \bibfield  {author} {\bibinfo {author} {\bibfnamefont {A.}~\bibnamefont
  {Mahmoud}}, \bibinfo {author} {\bibfnamefont {F.}~\bibnamefont {Ciubotaru}},
  \bibinfo {author} {\bibfnamefont {F.}~\bibnamefont {Vanderveken}}, \bibinfo
  {author} {\bibfnamefont {A.~V.}\ \bibnamefont {Chumak}}, \bibinfo {author}
  {\bibfnamefont {S.}~\bibnamefont {Hamdioui}}, \bibinfo {author}
  {\bibfnamefont {C.}~\bibnamefont {Adelmann}}, \ and\ \bibinfo {author}
  {\bibfnamefont {S.}~\bibnamefont {Cotofana}},\ }\href {\doibase
  10.1063/5.0019328} {\bibfield  {journal} {\bibinfo  {journal} {Journal of
  Applied Physics}\ }\textbf {\bibinfo {volume} {128}},\ \bibinfo {pages}
  {161101} (\bibinfo {year} {2020})},\ \Eprint
  {http://arxiv.org/abs/https://doi.org/10.1063/5.0019328}
  {https://doi.org/10.1063/5.0019328} \BibitemShut {NoStop}%
\bibitem [{\citenamefont {Barman}\ \emph {et~al.}(2021)\citenamefont {Barman}
  \emph {et~al.}}]{roadmap}%
  \BibitemOpen
  \bibfield  {author} {\bibinfo {author} {\bibfnamefont {A.}~\bibnamefont
  {Barman}} \emph {et~al.},\ }\href
  {http://iopscience.iop.org/article/10.1088/1361-648X/abec1a} {\bibfield
  {journal} {\bibinfo  {journal} {Journal of Physics: Condensed Matter}\ }
  (\bibinfo {year} {2021})}\BibitemShut {NoStop}%
\bibitem [{\citenamefont {{Mahmoud}}\ \emph
  {et~al.}(2021{\natexlab{a}})\citenamefont {{Mahmoud}}, \citenamefont
  {{Vanderveken}}, \citenamefont {{Adelmann}}, \citenamefont {{Ciubotaru}},
  \citenamefont {{Cotofana}},\ and\ \citenamefont {{Hamdioui}}}]{amahmoud1}%
  \BibitemOpen
  \bibfield  {author} {\bibinfo {author} {\bibfnamefont {A.~N.}\ \bibnamefont
  {{Mahmoud}}}, \bibinfo {author} {\bibfnamefont {F.}~\bibnamefont
  {{Vanderveken}}}, \bibinfo {author} {\bibfnamefont {C.}~\bibnamefont
  {{Adelmann}}}, \bibinfo {author} {\bibfnamefont {F.}~\bibnamefont
  {{Ciubotaru}}}, \bibinfo {author} {\bibfnamefont {S.}~\bibnamefont
  {{Cotofana}}}, \ and\ \bibinfo {author} {\bibfnamefont {S.}~\bibnamefont
  {{Hamdioui}}},\ }\href {\doibase 10.1109/TCSI.2020.3028050} {\bibfield
  {journal} {\bibinfo  {journal} {IEEE Transactions on Circuits and Systems I:
  Regular Papers}\ }\textbf {\bibinfo {volume} {68}},\ \bibinfo {pages} {536}
  (\bibinfo {year} {2021}{\natexlab{a}})}\BibitemShut {NoStop}%
\bibitem [{\citenamefont {{Mahmoud}}\ \emph
  {et~al.}(2020{\natexlab{a}})\citenamefont {{Mahmoud}}, \citenamefont
  {{Vanderveken}}, \citenamefont {{Ciubotaru}}, \citenamefont {{Adelmann}},
  \citenamefont {{Cotofana}},\ and\ \citenamefont {{Hamdioui}}}]{parallelism}%
  \BibitemOpen
  \bibfield  {author} {\bibinfo {author} {\bibfnamefont {A.}~\bibnamefont
  {{Mahmoud}}}, \bibinfo {author} {\bibfnamefont {F.}~\bibnamefont
  {{Vanderveken}}}, \bibinfo {author} {\bibfnamefont {F.}~\bibnamefont
  {{Ciubotaru}}}, \bibinfo {author} {\bibfnamefont {C.}~\bibnamefont
  {{Adelmann}}}, \bibinfo {author} {\bibfnamefont {S.}~\bibnamefont
  {{Cotofana}}}, \ and\ \bibinfo {author} {\bibfnamefont {S.}~\bibnamefont
  {{Hamdioui}}},\ }in\ \href@noop {} {\emph {\bibinfo {booktitle} {2020 Design,
  Automation Test in Europe Conference Exhibition (DATE)}}}\ (\bibinfo {year}
  {2020})\ pp.\ \bibinfo {pages} {642--645}\BibitemShut {NoStop}%
\bibitem [{\citenamefont {{Mahmoud}}\ \emph
  {et~al.}(2021{\natexlab{b}})\citenamefont {{Mahmoud}}, \citenamefont
  {{Vanderveken}}, \citenamefont {{Adelmann}}, \citenamefont {{Ciubotaru}},
  \citenamefont {{Hamdioui}},\ and\ \citenamefont {{Cotofana}}}]{parallelism1}%
  \BibitemOpen
  \bibfield  {author} {\bibinfo {author} {\bibfnamefont {A.~N.}\ \bibnamefont
  {{Mahmoud}}}, \bibinfo {author} {\bibfnamefont {F.}~\bibnamefont
  {{Vanderveken}}}, \bibinfo {author} {\bibfnamefont {C.}~\bibnamefont
  {{Adelmann}}}, \bibinfo {author} {\bibfnamefont {F.}~\bibnamefont
  {{Ciubotaru}}}, \bibinfo {author} {\bibfnamefont {S.}~\bibnamefont
  {{Hamdioui}}}, \ and\ \bibinfo {author} {\bibfnamefont {S.}~\bibnamefont
  {{Cotofana}}},\ }\href {\doibase 10.1109/TMAG.2021.3062022} {\bibfield
  {journal} {\bibinfo  {journal} {IEEE Transactions on Magnetics}\ ,\ \bibinfo
  {pages} {1}} (\bibinfo {year} {2021}{\natexlab{b}})}\BibitemShut {NoStop}%
\bibitem [{\citenamefont {{Mahmoud}}\ \emph
  {et~al.}(2020{\natexlab{b}})\citenamefont {{Mahmoud}}, \citenamefont
  {{Vanderveken}}, \citenamefont {{Adelmann}}, \citenamefont {{Ciubotaru}},
  \citenamefont {{Cotofana}},\ and\ \citenamefont {{Hamdioui}}}]{fanout10}%
  \BibitemOpen
  \bibfield  {author} {\bibinfo {author} {\bibfnamefont {A.}~\bibnamefont
  {{Mahmoud}}}, \bibinfo {author} {\bibfnamefont {F.}~\bibnamefont
  {{Vanderveken}}}, \bibinfo {author} {\bibfnamefont {C.}~\bibnamefont
  {{Adelmann}}}, \bibinfo {author} {\bibfnamefont {F.}~\bibnamefont
  {{Ciubotaru}}}, \bibinfo {author} {\bibfnamefont {S.}~\bibnamefont
  {{Cotofana}}}, \ and\ \bibinfo {author} {\bibfnamefont {S.}~\bibnamefont
  {{Hamdioui}}},\ }in\ \href@noop {} {\emph {\bibinfo {booktitle} {ISVLSI}}}\
  (\bibinfo {year} {2020})\ pp.\ \bibinfo {pages} {60--65}\BibitemShut
  {NoStop}%
\bibitem [{\citenamefont {{Mahmoud}}\ \emph
  {et~al.}(2020{\natexlab{c}})\citenamefont {{Mahmoud}}, \citenamefont
  {{Vanderveken}}, \citenamefont {{Adelmann}}, \citenamefont {{Ciubotaru}},
  \citenamefont {{Hamdioui}},\ and\ \citenamefont {{Cotofana}}}]{fanout11}%
  \BibitemOpen
  \bibfield  {author} {\bibinfo {author} {\bibfnamefont {A.}~\bibnamefont
  {{Mahmoud}}}, \bibinfo {author} {\bibfnamefont {F.}~\bibnamefont
  {{Vanderveken}}}, \bibinfo {author} {\bibfnamefont {C.}~\bibnamefont
  {{Adelmann}}}, \bibinfo {author} {\bibfnamefont {F.}~\bibnamefont
  {{Ciubotaru}}}, \bibinfo {author} {\bibfnamefont {S.}~\bibnamefont
  {{Hamdioui}}}, \ and\ \bibinfo {author} {\bibfnamefont {S.}~\bibnamefont
  {{Cotofana}}},\ }in\ \href {\doibase 10.1109/ICCD50377.2020.00062} {\emph
  {\bibinfo {booktitle} {2020 IEEE 38th International Conference on Computer
  Design (ICCD)}}}\ (\bibinfo {year} {2020})\ pp.\ \bibinfo {pages}
  {332--335}\BibitemShut {NoStop}%
\bibitem [{\citenamefont {Kostylev}\ \emph {et~al.}(2005)\citenamefont
  {Kostylev}, \citenamefont {Serga}, \citenamefont {Schneider}, \citenamefont
  {Leven},\ and\ \citenamefont {Hillebrands}}]{logic21}%
  \BibitemOpen
  \bibfield  {author} {\bibinfo {author} {\bibfnamefont {M.~P.}\ \bibnamefont
  {Kostylev}}, \bibinfo {author} {\bibfnamefont {A.~A.}\ \bibnamefont {Serga}},
  \bibinfo {author} {\bibfnamefont {T.}~\bibnamefont {Schneider}}, \bibinfo
  {author} {\bibfnamefont {B.}~\bibnamefont {Leven}}, \ and\ \bibinfo {author}
  {\bibfnamefont {B.}~\bibnamefont {Hillebrands}},\ }\href {\doibase
  10.1063/1.2089147} {\bibfield  {journal} {\bibinfo  {journal} {Applied
  Physics Letters}\ }\textbf {\bibinfo {volume} {87}},\ \bibinfo {pages}
  {153501} (\bibinfo {year} {2005})},\ \Eprint
  {http://arxiv.org/abs/https://doi.org/10.1063/1.2089147}
  {https://doi.org/10.1063/1.2089147} \BibitemShut {NoStop}%
\bibitem [{\citenamefont {Schneider}\ \emph {et~al.}(2008)\citenamefont
  {Schneider}, \citenamefont {Serga}, \citenamefont {Leven}, \citenamefont
  {Hillebrands}, \citenamefont {Stamps},\ and\ \citenamefont
  {Kostylev}}]{logic12}%
  \BibitemOpen
  \bibfield  {author} {\bibinfo {author} {\bibfnamefont {T.}~\bibnamefont
  {Schneider}}, \bibinfo {author} {\bibfnamefont {A.~A.}\ \bibnamefont
  {Serga}}, \bibinfo {author} {\bibfnamefont {B.}~\bibnamefont {Leven}},
  \bibinfo {author} {\bibfnamefont {B.}~\bibnamefont {Hillebrands}}, \bibinfo
  {author} {\bibfnamefont {R.~L.}\ \bibnamefont {Stamps}}, \ and\ \bibinfo
  {author} {\bibfnamefont {M.~P.}\ \bibnamefont {Kostylev}},\ }\href {\doibase
  10.1063/1.2834714} {\bibfield  {journal} {\bibinfo  {journal} {Applied
  Physics Letters}\ }\textbf {\bibinfo {volume} {92}},\ \bibinfo {pages}
  {022505} (\bibinfo {year} {2008})},\ \Eprint
  {http://arxiv.org/abs/https://doi.org/10.1063/1.2834714}
  {https://doi.org/10.1063/1.2834714} \BibitemShut {NoStop}%
\bibitem [{\citenamefont {Lee}\ and\ \citenamefont {Kim}(2008)}]{logic11}%
  \BibitemOpen
  \bibfield  {author} {\bibinfo {author} {\bibfnamefont {K.-S.}\ \bibnamefont
  {Lee}}\ and\ \bibinfo {author} {\bibfnamefont {S.-K.}\ \bibnamefont {Kim}},\
  }\href {\doibase 10.1063/1.2975235} {\bibfield  {journal} {\bibinfo
  {journal} {Journal of Applied Physics}\ }\textbf {\bibinfo {volume} {104}},\
  \bibinfo {pages} {053909} (\bibinfo {year} {2008})},\ \Eprint
  {http://arxiv.org/abs/https://doi.org/10.1063/1.2975235}
  {https://doi.org/10.1063/1.2975235} \BibitemShut {NoStop}%
\bibitem [{\citenamefont {Ustinova}\ \emph {et~al.}(2017)\citenamefont
  {Ustinova}, \citenamefont {Nikitin}, \citenamefont {Ustinov}, \citenamefont
  {Kalinikos},\ and\ \citenamefont {Lähderanta}}]{logic17}%
  \BibitemOpen
  \bibfield  {author} {\bibinfo {author} {\bibfnamefont {I.~A.}\ \bibnamefont
  {Ustinova}}, \bibinfo {author} {\bibfnamefont {A.~A.}\ \bibnamefont
  {Nikitin}}, \bibinfo {author} {\bibfnamefont {A.~B.}\ \bibnamefont
  {Ustinov}}, \bibinfo {author} {\bibfnamefont {B.~A.}\ \bibnamefont
  {Kalinikos}}, \ and\ \bibinfo {author} {\bibfnamefont {E.}~\bibnamefont
  {Lähderanta}},\ }in\ \href {\doibase 10.1109/EMCCompo.2017.7998091} {\emph
  {\bibinfo {booktitle} {2017 11th International Workshop on the
  Electromagnetic Compatibility of Integrated Circuits (EMCCompo)}}}\ (\bibinfo
  {year} {2017})\ pp.\ \bibinfo {pages} {104--107}\BibitemShut {NoStop}%
\bibitem [{\citenamefont {Mahmoud}\ \emph {et~al.}(2020)\citenamefont
  {Mahmoud}, \citenamefont {Vanderveken}, \citenamefont {Adelmann},
  \citenamefont {Ciubotaru}, \citenamefont {Hamdioui},\ and\ \citenamefont
  {Cotofana}}]{fanout}%
  \BibitemOpen
  \bibfield  {author} {\bibinfo {author} {\bibfnamefont {A.}~\bibnamefont
  {Mahmoud}}, \bibinfo {author} {\bibfnamefont {F.}~\bibnamefont
  {Vanderveken}}, \bibinfo {author} {\bibfnamefont {C.}~\bibnamefont
  {Adelmann}}, \bibinfo {author} {\bibfnamefont {F.}~\bibnamefont {Ciubotaru}},
  \bibinfo {author} {\bibfnamefont {S.}~\bibnamefont {Hamdioui}}, \ and\
  \bibinfo {author} {\bibfnamefont {S.}~\bibnamefont {Cotofana}},\ }\href
  {\doibase 10.1063/1.5134690} {\bibfield  {journal} {\bibinfo  {journal} {AIP
  Advances}\ }\textbf {\bibinfo {volume} {10}},\ \bibinfo {pages} {035119}
  (\bibinfo {year} {2020})},\ \Eprint
  {http://arxiv.org/abs/https://doi.org/10.1063/1.5134690}
  {https://doi.org/10.1063/1.5134690} \BibitemShut {NoStop}%
\bibitem [{\citenamefont {Fischer}\ \emph {et~al.}(2017)\citenamefont
  {Fischer}, \citenamefont {Kewenig}, \citenamefont {Bozhko}, \citenamefont
  {Serga}, \citenamefont {Syvorotka}, \citenamefont {Ciubotaru}, \citenamefont
  {Adelmann}, \citenamefont {Hillebrands},\ and\ \citenamefont
  {Chumak}}]{logic2}%
  \BibitemOpen
  \bibfield  {author} {\bibinfo {author} {\bibfnamefont {T.}~\bibnamefont
  {Fischer}}, \bibinfo {author} {\bibfnamefont {M.}~\bibnamefont {Kewenig}},
  \bibinfo {author} {\bibfnamefont {D.~A.}\ \bibnamefont {Bozhko}}, \bibinfo
  {author} {\bibfnamefont {A.~A.}\ \bibnamefont {Serga}}, \bibinfo {author}
  {\bibfnamefont {I.~I.}\ \bibnamefont {Syvorotka}}, \bibinfo {author}
  {\bibfnamefont {F.}~\bibnamefont {Ciubotaru}}, \bibinfo {author}
  {\bibfnamefont {C.}~\bibnamefont {Adelmann}}, \bibinfo {author}
  {\bibfnamefont {B.}~\bibnamefont {Hillebrands}}, \ and\ \bibinfo {author}
  {\bibfnamefont {A.~V.}\ \bibnamefont {Chumak}},\ }\href {\doibase
  10.1063/1.4979840} {\bibfield  {journal} {\bibinfo  {journal} {Applied
  Physics Letters}\ }\textbf {\bibinfo {volume} {110}},\ \bibinfo {pages}
  {152401} (\bibinfo {year} {2017})},\ \Eprint
  {http://arxiv.org/abs/https://doi.org/10.1063/1.4979840}
  {https://doi.org/10.1063/1.4979840} \BibitemShut {NoStop}%
\bibitem [{\citenamefont {Talmelli}\ \emph {et~al.}(2020)\citenamefont
  {Talmelli}, \citenamefont {Devolder}, \citenamefont {Tr{\"a}ger},
  \citenamefont {F{\"o}rster}, \citenamefont {Wintz}, \citenamefont {Weigand},
  \citenamefont {Stoll}, \citenamefont {Heyns}, \citenamefont {Sch{\"u}tz},
  \citenamefont {Radu}, \citenamefont {Gr{\"a}fe}, \citenamefont {Ciubotaru},\
  and\ \citenamefont {Adelmann}}]{logic3}%
  \BibitemOpen
  \bibfield  {author} {\bibinfo {author} {\bibfnamefont {G.}~\bibnamefont
  {Talmelli}}, \bibinfo {author} {\bibfnamefont {T.}~\bibnamefont {Devolder}},
  \bibinfo {author} {\bibfnamefont {N.}~\bibnamefont {Tr{\"a}ger}}, \bibinfo
  {author} {\bibfnamefont {J.}~\bibnamefont {F{\"o}rster}}, \bibinfo {author}
  {\bibfnamefont {S.}~\bibnamefont {Wintz}}, \bibinfo {author} {\bibfnamefont
  {M.}~\bibnamefont {Weigand}}, \bibinfo {author} {\bibfnamefont
  {H.}~\bibnamefont {Stoll}}, \bibinfo {author} {\bibfnamefont
  {M.}~\bibnamefont {Heyns}}, \bibinfo {author} {\bibfnamefont
  {G.}~\bibnamefont {Sch{\"u}tz}}, \bibinfo {author} {\bibfnamefont {I.~P.}\
  \bibnamefont {Radu}}, \bibinfo {author} {\bibfnamefont {J.}~\bibnamefont
  {Gr{\"a}fe}}, \bibinfo {author} {\bibfnamefont {F.}~\bibnamefont
  {Ciubotaru}}, \ and\ \bibinfo {author} {\bibfnamefont {C.}~\bibnamefont
  {Adelmann}},\ }\href {\doibase 10.1126/sciadv.abb4042} {\bibfield  {journal}
  {\bibinfo  {journal} {Science Advances}\ }\textbf {\bibinfo {volume} {6}}
  (\bibinfo {year} {2020}),\ 10.1126/sciadv.abb4042},\ \Eprint
  {http://arxiv.org/abs/https://advances.sciencemag.org/content/6/51/eabb4042.full.pdf}
  {https://advances.sciencemag.org/content/6/51/eabb4042.full.pdf} \BibitemShut
  {NoStop}%
\bibitem [{\citenamefont {{Ciubotaru}}\ \emph {et~al.}(2018)\citenamefont
  {{Ciubotaru}}, \citenamefont {{Talmelli}}, \citenamefont {{Devolder}},
  \citenamefont {{Zografos}} \emph {et~al.}}]{logic101}%
  \BibitemOpen
  \bibfield  {author} {\bibinfo {author} {\bibfnamefont {F.}~\bibnamefont
  {{Ciubotaru}}}, \bibinfo {author} {\bibfnamefont {G.}~\bibnamefont
  {{Talmelli}}}, \bibinfo {author} {\bibfnamefont {T.}~\bibnamefont
  {{Devolder}}}, \bibinfo {author} {\bibfnamefont {O.}~\bibnamefont
  {{Zografos}}},  \emph {et~al.},\ }in\ \href {\doibase
  10.1109/IEDM.2018.8614488} {\emph {\bibinfo {booktitle} {2018 IEEE
  International Electron Devices Meeting (IEDM)}}}\ (\bibinfo {year} {2018})\
  pp.\ \bibinfo {pages} {36.1.1--36.1.4}\BibitemShut {NoStop}%
\bibitem [{\citenamefont {Khitun}\ and\ \citenamefont {Wang}(2011)}]{logic1}%
  \BibitemOpen
  \bibfield  {author} {\bibinfo {author} {\bibfnamefont {A.}~\bibnamefont
  {Khitun}}\ and\ \bibinfo {author} {\bibfnamefont {K.~L.}\ \bibnamefont
  {Wang}},\ }\href {\doibase 10.1063/1.3609062} {\bibfield  {journal} {\bibinfo
   {journal} {Journal of Applied Physics}\ }\textbf {\bibinfo {volume} {110}},\
  \bibinfo {pages} {034306} (\bibinfo {year} {2011})},\ \Eprint
  {http://arxiv.org/abs/https://doi.org/10.1063/1.3609062}
  {https://doi.org/10.1063/1.3609062} \BibitemShut {NoStop}%
\bibitem [{\citenamefont {Gertz}\ \emph {et~al.}(2015)\citenamefont {Gertz},
  \citenamefont {Kozhevnikov}, \citenamefont {Filimonov},\ and\ \citenamefont
  {Khitun}}]{memory3}%
  \BibitemOpen
  \bibfield  {author} {\bibinfo {author} {\bibfnamefont {F.}~\bibnamefont
  {Gertz}}, \bibinfo {author} {\bibfnamefont {A.}~\bibnamefont {Kozhevnikov}},
  \bibinfo {author} {\bibfnamefont {Y.}~\bibnamefont {Filimonov}}, \ and\
  \bibinfo {author} {\bibfnamefont {A.}~\bibnamefont {Khitun}},\ }\href
  {\doibase 10.1109/TMAG.2014.2362723} {\bibfield  {journal} {\bibinfo
  {journal} {IEEE Transactions on Magnetics}\ }\textbf {\bibinfo {volume}
  {51}},\ \bibinfo {pages} {1} (\bibinfo {year} {2015})}\BibitemShut {NoStop}%
\bibitem [{\citenamefont {Wang}\ \emph {et~al.}(2020)\citenamefont {Wang},
  \citenamefont {Kewenig}, \citenamefont {Schneider}, \citenamefont {Verba},
  \citenamefont {Kohl}, \citenamefont {Heinz}, \citenamefont {Geilen},
  \citenamefont {Mohseni}, \citenamefont {L{\"a}gel}, \citenamefont {Ciubotaru}
  \emph {et~al.}}]{wang2020magnonic}%
  \BibitemOpen
  \bibfield  {author} {\bibinfo {author} {\bibfnamefont {Q.}~\bibnamefont
  {Wang}}, \bibinfo {author} {\bibfnamefont {M.}~\bibnamefont {Kewenig}},
  \bibinfo {author} {\bibfnamefont {M.}~\bibnamefont {Schneider}}, \bibinfo
  {author} {\bibfnamefont {R.}~\bibnamefont {Verba}}, \bibinfo {author}
  {\bibfnamefont {F.}~\bibnamefont {Kohl}}, \bibinfo {author} {\bibfnamefont
  {B.}~\bibnamefont {Heinz}}, \bibinfo {author} {\bibfnamefont
  {M.}~\bibnamefont {Geilen}}, \bibinfo {author} {\bibfnamefont
  {M.}~\bibnamefont {Mohseni}}, \bibinfo {author} {\bibfnamefont
  {B.}~\bibnamefont {L{\"a}gel}}, \bibinfo {author} {\bibfnamefont
  {F.}~\bibnamefont {Ciubotaru}},  \emph {et~al.},\ }\href@noop {} {\bibfield
  {journal} {\bibinfo  {journal} {Nature Electronics}\ ,\ \bibinfo {pages} {1}}
  (\bibinfo {year} {2020})}\BibitemShut {NoStop}%
\bibitem [{\citenamefont {Mittal}(2016)}]{applications}%
  \BibitemOpen
  \bibfield  {author} {\bibinfo {author} {\bibfnamefont {S.}~\bibnamefont
  {Mittal}},\ }\href {\doibase 10.1145/2893356} {\bibfield  {journal} {\bibinfo
   {journal} {ACM Comput. Surv.}\ }\textbf {\bibinfo {volume} {48}} (\bibinfo
  {year} {2016}),\ 10.1145/2893356}\BibitemShut {NoStop}%
\bibitem [{\citenamefont {{Angizi}}\ \emph {et~al.}(2018)\citenamefont
  {{Angizi}}, \citenamefont {{Jiang}}, \citenamefont {{DeMara}}, \citenamefont
  {{Han}},\ and\ \citenamefont {{Fan}}}]{SPIN}%
  \BibitemOpen
  \bibfield  {author} {\bibinfo {author} {\bibfnamefont {S.}~\bibnamefont
  {{Angizi}}}, \bibinfo {author} {\bibfnamefont {H.}~\bibnamefont {{Jiang}}},
  \bibinfo {author} {\bibfnamefont {R.~F.}\ \bibnamefont {{DeMara}}}, \bibinfo
  {author} {\bibfnamefont {J.}~\bibnamefont {{Han}}}, \ and\ \bibinfo {author}
  {\bibfnamefont {D.}~\bibnamefont {{Fan}}},\ }\href@noop {} {\bibfield
  {journal} {\bibinfo  {journal} {IEEE Transactions on Nanotechnology}\
  }\textbf {\bibinfo {volume} {17}},\ \bibinfo {pages} {795} (\bibinfo {year}
  {2018})}\BibitemShut {NoStop}%
\bibitem [{\citenamefont {Devolder}\ \emph {et~al.}(2016)\citenamefont
  {Devolder}, \citenamefont {Kim}, \citenamefont {Garcia-Sanchez},
  \citenamefont {Swerts}, \citenamefont {Kim}, \citenamefont {Couet},
  \citenamefont {Kar},\ and\ \citenamefont {Furnemont}}]{parameters}%
  \BibitemOpen
  \bibfield  {author} {\bibinfo {author} {\bibfnamefont {T.}~\bibnamefont
  {Devolder}}, \bibinfo {author} {\bibfnamefont {J.-V.}\ \bibnamefont {Kim}},
  \bibinfo {author} {\bibfnamefont {F.}~\bibnamefont {Garcia-Sanchez}},
  \bibinfo {author} {\bibfnamefont {J.}~\bibnamefont {Swerts}}, \bibinfo
  {author} {\bibfnamefont {W.}~\bibnamefont {Kim}}, \bibinfo {author}
  {\bibfnamefont {S.}~\bibnamefont {Couet}}, \bibinfo {author} {\bibfnamefont
  {G.}~\bibnamefont {Kar}}, \ and\ \bibinfo {author} {\bibfnamefont
  {A.}~\bibnamefont {Furnemont}},\ }\href {\doibase 10.1103/PhysRevB.93.024420}
  {\bibfield  {journal} {\bibinfo  {journal} {Phys. Rev. B}\ }\textbf {\bibinfo
  {volume} {93}},\ \bibinfo {pages} {024420} (\bibinfo {year}
  {2016})}\BibitemShut {NoStop}%
\bibitem [{\citenamefont {Vansteenkiste}\ \emph {et~al.}(2014)\citenamefont
  {Vansteenkiste}, \citenamefont {Leliaert}, \citenamefont {Dvornik},
  \citenamefont {Helsen}, \citenamefont {Garcia-Sanchez},\ and\ \citenamefont
  {Van~Waeyenberge}}]{mumax}%
  \BibitemOpen
  \bibfield  {author} {\bibinfo {author} {\bibfnamefont {A.}~\bibnamefont
  {Vansteenkiste}}, \bibinfo {author} {\bibfnamefont {J.}~\bibnamefont
  {Leliaert}}, \bibinfo {author} {\bibfnamefont {M.}~\bibnamefont {Dvornik}},
  \bibinfo {author} {\bibfnamefont {M.}~\bibnamefont {Helsen}}, \bibinfo
  {author} {\bibfnamefont {F.}~\bibnamefont {Garcia-Sanchez}}, \ and\ \bibinfo
  {author} {\bibfnamefont {B.}~\bibnamefont {Van~Waeyenberge}},\ }\href
  {\doibase 10.1063/1.4899186} {\bibfield  {journal} {\bibinfo  {journal} {AIP
  Advances}\ }\textbf {\bibinfo {volume} {4}},\ \bibinfo {pages} {107133}
  (\bibinfo {year} {2014})},\ \Eprint
  {http://arxiv.org/abs/https://doi.org/10.1063/1.4899186}
  {https://doi.org/10.1063/1.4899186} \BibitemShut {NoStop}%
\bibitem [{\citenamefont {Mahmoud}\ \emph {et~al.}(2021)\citenamefont
  {Mahmoud}, \citenamefont {Vanderveken}, \citenamefont {Ciubotaru},
  \citenamefont {Adelmann}, \citenamefont {Cotofana},\ and\ \citenamefont
  {Hamdioui}}]{SW2}%
  \BibitemOpen
  \bibfield  {author} {\bibinfo {author} {\bibfnamefont {A.}~\bibnamefont
  {Mahmoud}}, \bibinfo {author} {\bibfnamefont {F.}~\bibnamefont
  {Vanderveken}}, \bibinfo {author} {\bibfnamefont {F.}~\bibnamefont
  {Ciubotaru}}, \bibinfo {author} {\bibfnamefont {C.}~\bibnamefont {Adelmann}},
  \bibinfo {author} {\bibfnamefont {S.}~\bibnamefont {Cotofana}}, \ and\
  \bibinfo {author} {\bibfnamefont {S.}~\bibnamefont {Hamdioui}},\ }\href@noop
  {} {\enquote {\bibinfo {title} {Spin wave based full adder},}\ } (\bibinfo
  {year} {2021}),\ \Eprint {http://arxiv.org/abs/2102.08108} {arXiv:2102.08108
  [cond-mat.mes-hall]} \BibitemShut {NoStop}%
\bibitem [{\citenamefont {{Canan}}\ \emph {et~al.}(2019)\citenamefont
  {{Canan}}, \citenamefont {{Kaya}}, \citenamefont {{Karanth}},\ and\
  \citenamefont {{Louri}}}]{CMOS1}%
  \BibitemOpen
  \bibfield  {author} {\bibinfo {author} {\bibfnamefont {T.~F.}\ \bibnamefont
  {{Canan}}}, \bibinfo {author} {\bibfnamefont {S.}~\bibnamefont {{Kaya}}},
  \bibinfo {author} {\bibfnamefont {A.}~\bibnamefont {{Karanth}}}, \ and\
  \bibinfo {author} {\bibfnamefont {A.}~\bibnamefont {{Louri}}},\ }\href
  {\doibase 10.1109/JXCDC.2019.2962494} {\bibfield  {journal} {\bibinfo
  {journal} {IEEE Journal on Exploratory Solid-State Computational Devices and
  Circuits}\ }\textbf {\bibinfo {volume} {5}},\ \bibinfo {pages} {94} (\bibinfo
  {year} {2019})}\BibitemShut {NoStop}%
\bibitem [{\citenamefont {{Roohi}}\ \emph {et~al.}(2017)\citenamefont
  {{Roohi}}, \citenamefont {{Zand}}, \citenamefont {{Fan}},\ and\ \citenamefont
  {{DeMara}}}]{SHE2}%
  \BibitemOpen
  \bibfield  {author} {\bibinfo {author} {\bibfnamefont {A.}~\bibnamefont
  {{Roohi}}}, \bibinfo {author} {\bibfnamefont {R.}~\bibnamefont {{Zand}}},
  \bibinfo {author} {\bibfnamefont {D.}~\bibnamefont {{Fan}}}, \ and\ \bibinfo
  {author} {\bibfnamefont {R.~F.}\ \bibnamefont {{DeMara}}},\ }\href@noop {}
  {\bibfield  {journal} {\bibinfo  {journal} {IEEE Transactions on
  Computer-Aided Design of Integrated Circuits and Systems}\ }\textbf {\bibinfo
  {volume} {36}},\ \bibinfo {pages} {2134} (\bibinfo {year}
  {2017})}\BibitemShut {NoStop}%
\bibitem [{\citenamefont {{Roohi}}, \citenamefont {{Zand}},\ and\ \citenamefont
  {{DeMara}}(2016)}]{DWM}%
  \BibitemOpen
  \bibfield  {author} {\bibinfo {author} {\bibfnamefont {A.}~\bibnamefont
  {{Roohi}}}, \bibinfo {author} {\bibfnamefont {R.}~\bibnamefont {{Zand}}}, \
  and\ \bibinfo {author} {\bibfnamefont {R.~F.}\ \bibnamefont {{DeMara}}},\
  }\href@noop {} {\bibfield  {journal} {\bibinfo  {journal} {IEEE Transactions
  on Magnetics}\ }\textbf {\bibinfo {volume} {52}},\ \bibinfo {pages} {1}
  (\bibinfo {year} {2016})}\BibitemShut {NoStop}%
\bibitem [{\citenamefont {{Gupta}}\ \emph {et~al.}(2011)\citenamefont
  {{Gupta}}, \citenamefont {{Mohapatra}}, \citenamefont {{Park}}, \citenamefont
  {{Raghunathan}},\ and\ \citenamefont {{Roy}}}]{approxCMOS1}%
  \BibitemOpen
  \bibfield  {author} {\bibinfo {author} {\bibfnamefont {V.}~\bibnamefont
  {{Gupta}}}, \bibinfo {author} {\bibfnamefont {D.}~\bibnamefont
  {{Mohapatra}}}, \bibinfo {author} {\bibfnamefont {S.~P.}\ \bibnamefont
  {{Park}}}, \bibinfo {author} {\bibfnamefont {A.}~\bibnamefont
  {{Raghunathan}}}, \ and\ \bibinfo {author} {\bibfnamefont {K.}~\bibnamefont
  {{Roy}}},\ }in\ \href {\doibase 10.1109/ISLPED.2011.5993675} {\emph {\bibinfo
  {booktitle} {IEEE/ACM International Symposium on Low Power Electronics and
  Design}}}\ (\bibinfo {year} {2011})\ pp.\ \bibinfo {pages}
  {409--414}\BibitemShut {NoStop}%
\bibitem [{\citenamefont {{Cai}}\ \emph {et~al.}(2017)\citenamefont {{Cai}},
  \citenamefont {{Wang}}, \citenamefont {{De Barros Naviner}},\ and\
  \citenamefont {{Zhao}}}]{MTJ2}%
  \BibitemOpen
  \bibfield  {author} {\bibinfo {author} {\bibfnamefont {H.}~\bibnamefont
  {{Cai}}}, \bibinfo {author} {\bibfnamefont {Y.}~\bibnamefont {{Wang}}},
  \bibinfo {author} {\bibfnamefont {L.~A.}\ \bibnamefont {{De Barros
  Naviner}}}, \ and\ \bibinfo {author} {\bibfnamefont {W.}~\bibnamefont
  {{Zhao}}},\ }\href@noop {} {\bibfield  {journal} {\bibinfo  {journal} {IEEE
  Transactions on Circuits and Systems I: Regular Papers}\ }\textbf {\bibinfo
  {volume} {64}},\ \bibinfo {pages} {847} (\bibinfo {year} {2017})}\BibitemShut
  {NoStop}%
\bibitem [{\citenamefont {{Zografos}}\ \emph {et~al.}(2015)\citenamefont
  {{Zografos}}, \citenamefont {{Sorée}}, \citenamefont {{Vaysset}},
  \citenamefont {{Cosemans}}, \citenamefont {{Amarù}}, \citenamefont
  {{Gaillardon}}, \citenamefont {{De Micheli}}, \citenamefont {{Lauwereins}},
  \citenamefont {{Sayan}}, \citenamefont {{Raghavan}}, \citenamefont {{Radu}},\
  and\ \citenamefont {{Thean}}}]{Excitation_table_ref16}%
  \BibitemOpen
  \bibfield  {author} {\bibinfo {author} {\bibfnamefont {O.}~\bibnamefont
  {{Zografos}}}, \bibinfo {author} {\bibfnamefont {B.}~\bibnamefont
  {{Sorée}}}, \bibinfo {author} {\bibfnamefont {A.}~\bibnamefont {{Vaysset}}},
  \bibinfo {author} {\bibfnamefont {S.}~\bibnamefont {{Cosemans}}}, \bibinfo
  {author} {\bibfnamefont {L.}~\bibnamefont {{Amarù}}}, \bibinfo {author}
  {\bibfnamefont {P.}~\bibnamefont {{Gaillardon}}}, \bibinfo {author}
  {\bibfnamefont {G.}~\bibnamefont {{De Micheli}}}, \bibinfo {author}
  {\bibfnamefont {R.}~\bibnamefont {{Lauwereins}}}, \bibinfo {author}
  {\bibfnamefont {S.}~\bibnamefont {{Sayan}}}, \bibinfo {author} {\bibfnamefont
  {P.}~\bibnamefont {{Raghavan}}}, \bibinfo {author} {\bibfnamefont {I.~P.}\
  \bibnamefont {{Radu}}}, \ and\ \bibinfo {author} {\bibfnamefont
  {A.}~\bibnamefont {{Thean}}},\ }in\ \href {\doibase
  10.1109/NANO.2015.7388699} {\emph {\bibinfo {booktitle} {2015 IEEE 15th
  International Conference on Nanotechnology (IEEE-NANO)}}}\ (\bibinfo {year}
  {2015})\ pp.\ \bibinfo {pages} {686--689}\BibitemShut {NoStop}%
\bibitem [{\citenamefont {{Kulkarni}}, \citenamefont {{Gupta}},\ and\
  \citenamefont {{Ercegovac}}(2011)}]{approxCMOS2}%
  \BibitemOpen
  \bibfield  {author} {\bibinfo {author} {\bibfnamefont {P.}~\bibnamefont
  {{Kulkarni}}}, \bibinfo {author} {\bibfnamefont {P.}~\bibnamefont {{Gupta}}},
  \ and\ \bibinfo {author} {\bibfnamefont {M.}~\bibnamefont {{Ercegovac}}},\
  }in\ \href {\doibase 10.1109/VLSID.2011.51} {\emph {\bibinfo {booktitle}
  {2011 24th Internatioal Conference on VLSI Design}}}\ (\bibinfo {year}
  {2011})\ pp.\ \bibinfo {pages} {346--351}\BibitemShut {NoStop}%
\bibitem [{\citenamefont {{Chen}}\ \emph {et~al.}(2013)\citenamefont {{Chen}},
  \citenamefont {{Sangai}}, \citenamefont {{Gholipour}},\ and\ \citenamefont
  {{Chen}}}]{16nmCMOS}%
  \BibitemOpen
  \bibfield  {author} {\bibinfo {author} {\bibfnamefont {Y.}~\bibnamefont
  {{Chen}}}, \bibinfo {author} {\bibfnamefont {A.}~\bibnamefont {{Sangai}}},
  \bibinfo {author} {\bibfnamefont {M.}~\bibnamefont {{Gholipour}}}, \ and\
  \bibinfo {author} {\bibfnamefont {D.}~\bibnamefont {{Chen}}},\ }in\ \href
  {\doibase 10.1109/NanoArch.2013.6623049} {\emph {\bibinfo {booktitle} {2013
  IEEE/ACM International Symposium on Nanoscale Architectures (NANOARCH)}}}\
  (\bibinfo {year} {2013})\ pp.\ \bibinfo {pages} {82--88}\BibitemShut
  {NoStop}%
\bibitem [{\citenamefont {Wang}\ \emph {et~al.}(2018)\citenamefont {Wang},
  \citenamefont {Pirro}, \citenamefont {Verba}, \citenamefont {Slavin},
  \citenamefont {Hillebrands},\ and\ \citenamefont {Chumak}}]{DC}%
  \BibitemOpen
  \bibfield  {author} {\bibinfo {author} {\bibfnamefont {Q.}~\bibnamefont
  {Wang}}, \bibinfo {author} {\bibfnamefont {P.}~\bibnamefont {Pirro}},
  \bibinfo {author} {\bibfnamefont {R.}~\bibnamefont {Verba}}, \bibinfo
  {author} {\bibfnamefont {A.}~\bibnamefont {Slavin}}, \bibinfo {author}
  {\bibfnamefont {B.}~\bibnamefont {Hillebrands}}, \ and\ \bibinfo {author}
  {\bibfnamefont {A.~V.}\ \bibnamefont {Chumak}},\ }\href {\doibase
  10.1126/sciadv.1701517} {\bibfield  {journal} {\bibinfo  {journal} {Science
  Advances}\ }\textbf {\bibinfo {volume} {4}} (\bibinfo {year} {2018}),\
  10.1126/sciadv.1701517},\ \Eprint
  {http://arxiv.org/abs/https://advances.sciencemag.org/content/4/1/e1701517.full.pdf}
  {https://advances.sciencemag.org/content/4/1/e1701517.full.pdf} \BibitemShut
  {NoStop}%
\end{thebibliography}%

\end{document}